\documentclass[%
reprint,
amsmath,amssymb,
aps,
]{revtex4-2}

\usepackage{graphicx}% Include figure files
\usepackage{dcolumn}% Align table columns on decimal point
\usepackage{bm}% bold math

\usepackage[tight-spacing = false, per-mode=repeated-symbol]{siunitx} % For displaying SI units and numbers correctly

\begin{document}

\bibliographystyle{apsrev4-2}

\preprint{APS/123-QED}

\title{Broken Symmetries in Microfluidic Pillar Arrays are Reflected in a Flowing DNA Solution across Multiple Length Scales}% Force line breaks with \\

\author{Jason P. Beech*}
\author{Oskar E. Ström*}%
 \author{Jonas O. Tegenfeldt†}%
\affiliation{%
 Division of Solid State Physics, Department of Physics and NanoLund, Lund University, P.O. Box 118, 22100 Lund, Sweden}%

\date{\today}%

\begin{abstract}% max 600 characters in the abstract according to the instructions (currently 601 characters incl space)
Unlike Newtonian fluids, viscoelastic fluids may break time-reversal symmetry at low Reynolds numbers resulting in elastic turbulence. Furthermore, under some conditions, instead of the chaotic turbulence, large-scale regular waves form, as has been shown for DNA flowing in microfluidic pillar arrays. We here demonstrate how the symmetry of the individual pillars influences the symmetry of these waves, thereby contributing to the understanding of the origin of the waves and opening up for better control of the waves with relevance to applications such as microfluidic sorting and mixing. The onset of waves occurs at different Deborah numbers for flow in different directions through the same array. Because the onset of waves leads to an increase in flow rate for a given driving pressure, we observe an increase in diodicity within this range.
\\ * Authors contributed equally.\\$\dagger$ Corresponding author. jonas.tegenfeldt@ftf.lth.se\\
\end{abstract}

\maketitle

Viscoelastic effects in the flows of polymer solutions through and around structures at the microscale are believed to have practical importance in diverse areas ranging from the extraction of oil from mineral matrices [\onlinecite{Sorbie1991_book}] to the understanding of the transport of biological fluids, both in their native environments [\onlinecite{Carrel2018}] and through novel types of microfluidics devices [\onlinecite{Groisman2011}]. An example of the latter are microfluidics devices for the separation of DNA molecules by length that use deterministic lateral displacement (DLD)  [\onlinecite{Hochstetter2020, Huang2004, Chen2015, Wunsch2019, Wunsch2021, strom2022}] as well those relying on pulsed electrical fields combined with a pillar array arranged in an hexagonal pattern [\onlinecite{bakajin_hex2001, huang_hex2002}]. While the underlying mechanisms are entirely different, both are based on arrays of pillars through which the molecules follow size dependent trajectories. Since solutions of DNA are viscoelastic fluids, an understanding of their behavior in pillar arrays is essential for the development and improvement in functionality of these devices. Indeed, during recent efforts to improve the separation of long ($>$ 20 kbp) DNA fragments in DLD devices [\onlinecite{strom2022}] we discovered that DNA can assemble into ordered waves of high concentration and molecular stretching and what is more that the local flow direction inside these waves varies from the bulk flow direction, an effect which both increased and decreased the quality of separation dependent on the specific conditions. We have subsequently shown that the formation of these waves depends to large degree on the viscoelastic properties of the fluid and the order of the pillar array, with large scale flow patterns or waves entirely absent in arrays where the position of each pillar has been randomly perturbed [\onlinecite{strom_waves2022}].

Previous research found instabilities in viscoelastic flows through arrays of symmetric [\onlinecite{kawale2017}] and non-symmetric (triangular) [\onlinecite{kawale2019}] pillars. Triangular pillars [\onlinecite{Loutherback2009, Loutherback2010}] have been used to decrease both the critical size and propensity for clogging in DLD devices. Here we show that by breaking the symmetry of the pillars, a microscopic property of the array, and doing so both parallel with and perpendicular to the flow direction, we observe symmetry breaking in the flow behavior of the DNA solution at macroscopic scales.

\begin{figure*}[htbp]
\includegraphics[width=165mm]{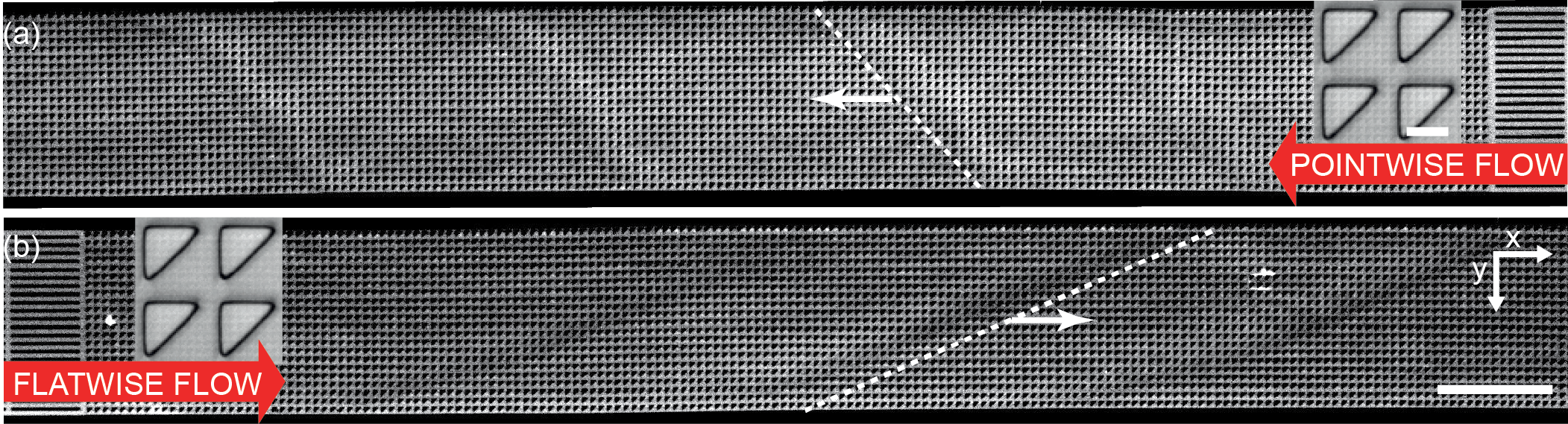}
\caption{\label{fig:snapshots_array}Low-magnification (2$\times$) snapshots of fluorescent DNA moving through an array of triangular pillars in the pointwise ($u\approx$ \SI{8.7}{\milli\meter\per\second}, $De \approx 350$) (a) and the flatwise ($u\approx$ \SI{10}{\milli\meter\per\second}, $De \approx 400$) (b) directions. For each flow direction, one wave is highlighted with dashed white line. The dynamics of the waves is clearly seen in Supplemental Movie S1 [\onlinecite{SI}]. The intensity and brightness are the same for all figure panels and set to enhance the visibility of the waves (as for all subsequent figures). Scale bars are 500~$\mu$m for the low-magnification snapshots and 20 $\mu$m in the inset.}
\end{figure*}

Our device consists of a straight channel, 0.8~mm wide, 8~mm long, and 111.5 µm deep, containing an array (22~$\times$~222) of right angled isosceles triangular pillars with a pitch of 36~µm. The legs of the triangles are 28~µm, the gaps between are 8 µm. The pillars are oriented such that flow in one direction impinges on a leg (which we will call flatwise flow), and in the other direction on an acute vertice (which we will call pointwise flow).
The devices were fabricated using standard replica molding techniques. 
$\lambda$ phage DNA was stained with the bisintercalating dye YOYO-1 at a DNA basepair to dye ratio of 200:1 in a 5$\times$ Tris EDTA (TE)-buffer.
The flow was generated by applying nitrogen gas overpressure and the flow rate was measured using a flow sensor. The flowing DNA was imaged using standard epifluorescence microscopy. Any confounding patterns caused by diffraction/reflection in the array were characterised in devices without DNA.
Full details of the materials and methods are described in the Supplemental Material [\onlinecite{SI}].

\begin{figure*}[!htbp]
\includegraphics[width=178mm]{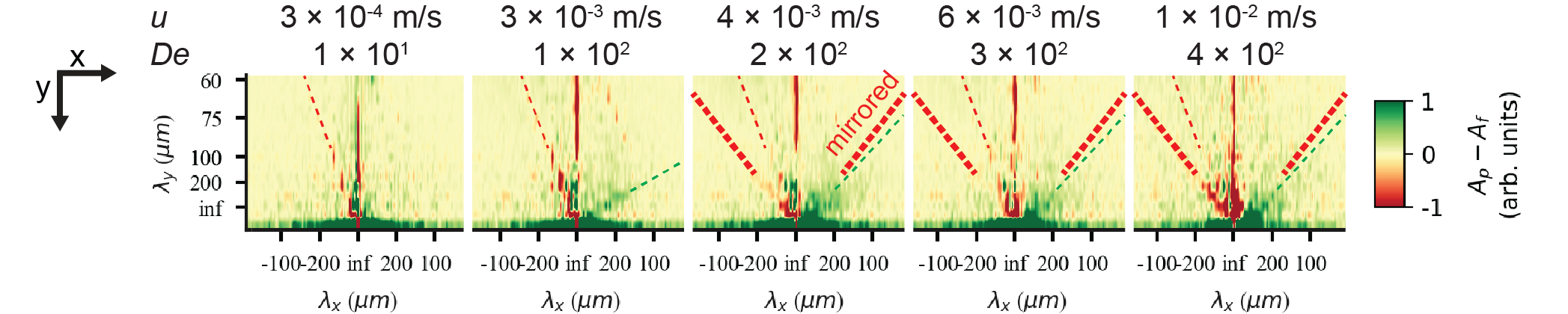}
\caption{\label{fig:fourier_spectra} Low frequency region of Fourier amplitude spectra difference for pointwise minus flatwise flows. Before subtraction, the mean spectra for all the frames of each videograph were computed. The data is based on fluorescence videos captured at a low magnification (2$\times$~objective) of various lengths in the range of 100 to 300 s. Guides for the eye are added as follows. The thin red dashed line (all panels) corresponds to an optical artefact (see section 5 in the Supplemental Material [\onlinecite{SI}]). The thick red dashed line corresponds to waves for flatwise flow (mirrored for comparison with pointwise waves). The dashed green line corresponds to waves for pointwise flow. Note that pointwise waves occur at lower \textit{De} ($De=100$ for pointwise waves compared to $De=200$ for flatwise waves). See Supplemental Material, section 4 [\onlinecite{SI}] for details.} 
\end{figure*}

\begin{figure*}[!htbp]
\includegraphics[width=172mm]{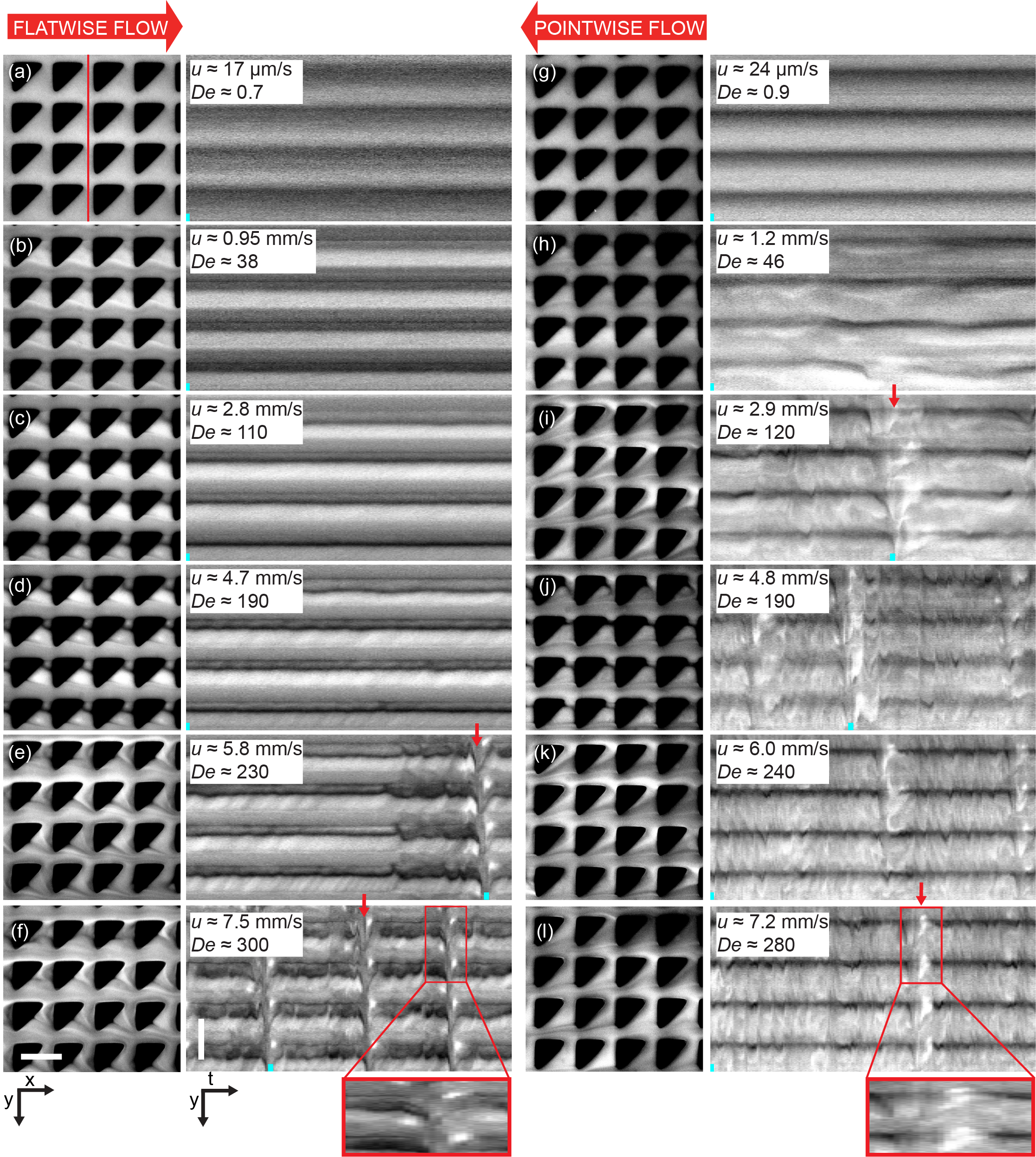}
\caption{\label{fig:high_res_micrographs_and_kymographs}Observed at the microscale (60$\times$ magnification), the flow behavior differs significantly between the two flow directions. The pairs of images in each panel consist of micrographs ($x$,$y$) and kymographs ($y$,$t$) taken along the vertical line shown in the first panel. The small bars found at the bottom of each kymograph indicate the relative time points of the micrograph in each pair. The arrows highlight a selection of the time points when waves occur. Each kymograph spans a total of 7 s. To illustrate the evolution of the high-concentration DNA blobs, region of interest spanning 0.9 s are indicated in (f) and (l) and expended in the time dimension at the very bottom. See Supplemental Movie S2 [\onlinecite{SI}] for videos corresponding to the data of all the panels. Scale bars are 20 $\mu$m.}
\end{figure*}

\begin{figure*}[!htbp]
\includegraphics[width=178mm]{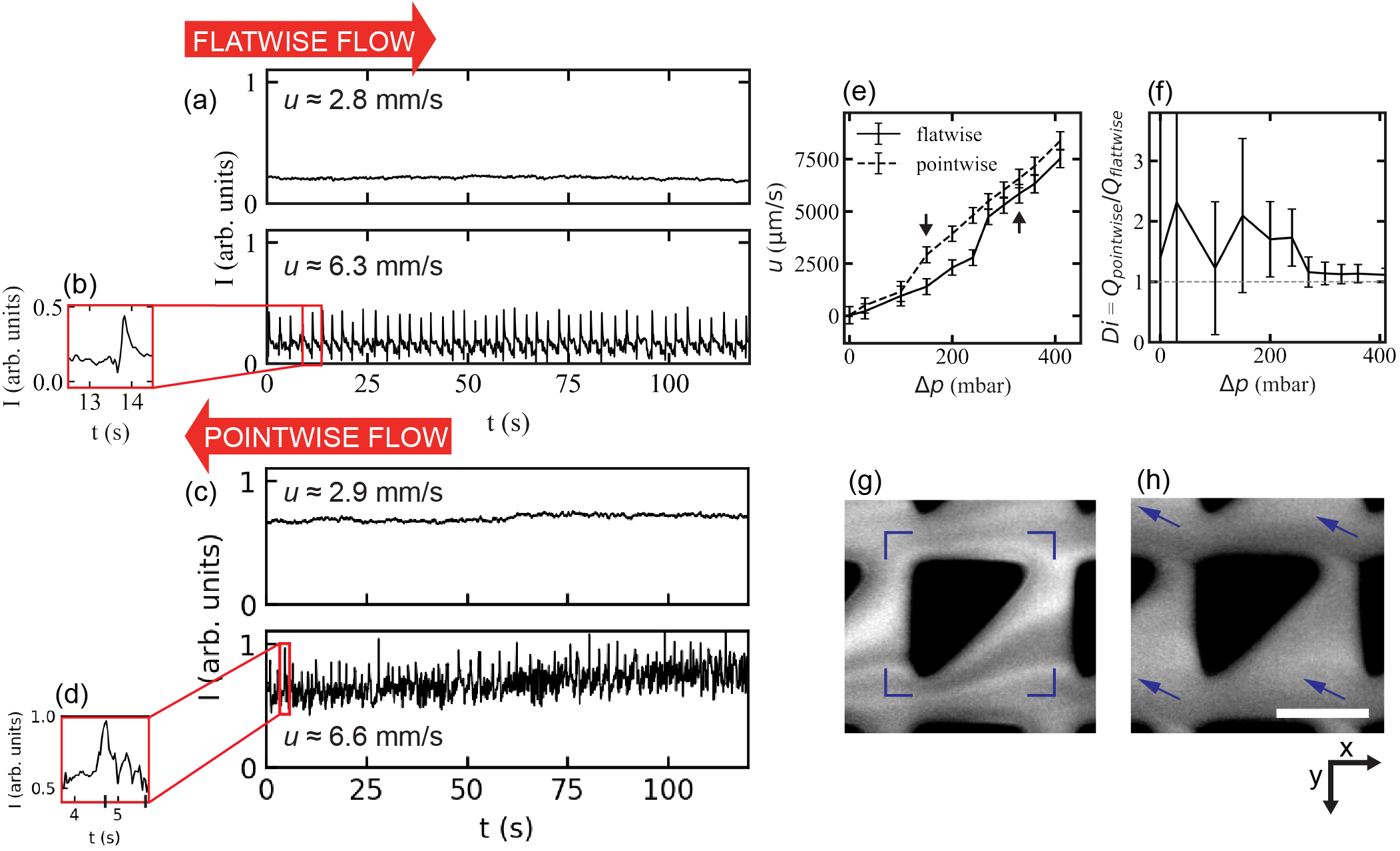}
\caption{\label{fig:high_res_time_signal_analysis} Microscopic concentration fluctuations (a-d) as measured around one pillar as indicated in (f). (b) and (d) provide detailed views of the peaks from the plots (a) and (c) marked with red rectangles. The vertical bars crossing the time axis in (d) corresponds to the timepoints for which (g) and (h) were acquired. Flow characteristics around one pillar in pointwise flow (g and h) [20$\times$ magnification]. The overall dynamics can be observed in Supplemental Movies S2 and S3 [\onlinecite{SI}]. Scale bar is 20 $\mu$m. (e) Mean flow velocity dependence of applied pressure and (f) Diodicity, $Di=Q_{\rm pointwise}/Q_{\rm flatwise}$. The arrows in (e) mark the earliest wave observations in pointwise (150 mbar, left arrow) and flatwise (330 mbar, right arrow). The error bars in (e) represent one standard deviation and those in (f) represent the fractional standard deviations added in quadrature. }
\end{figure*}

As a solution of $\lambda$ phage DNA at \SI{400}{\micro\gram\per\milli\litre} is forced to flow through the pillar array at various applied pressures and in the two different directions, several phenomena are observed with large qualitative differences immediately apparent. Figs.~\ref{fig:snapshots_array} and \ref{fig:fourier_spectra} describe the flow behavior in the two flow directions at the device scale and Figs.~\ref{fig:high_res_micrographs_and_kymographs} and \ref{fig:high_res_time_signal_analysis} describe the flow behavior at the scale of the individual pillars. For low-resolution micrographs for a larger range of flow velocities, see Supplemental Material Figs.~S2 and S3 [\onlinecite{SI}].

To be able to relate our observations to the prevailing flow conditions we derive the flow velocities from the measured volumetric flow rates, $Q$, 0.01\textendash\SI{8}{\nano\litre\per\second}, dividing by the total cross-sectional area of the smallest gaps between the pillars, $A$. We estimate the Reynolds number, \textit{Re} [Eq.~(\ref{eq:Re})], for all experiments to be in the range $9.3~\times~10^{-5}$ to $6~\times~10^{-2}$, such that we can neglect any inertial contributions:
\begin{eqnarray}
\label{eq:Re}
Re = \frac{\rho u_{\rm max} G}{\mu},
\end{eqnarray}

The Deborah number is evaluated using equation 2:
\begin{eqnarray}
\label{eq:De}
De = \frac{\tau u}{L}.
\end{eqnarray}
Here $\tau$ = 1.43 s is the relaxation time for $\lambda$ DNA (measurement detailed in [\onlinecite{strom_waves2022}]), and $L=~\SI{36e-6}{\meter}$ is the array pitch.
$u$ is the mean flow velocity between pillars (half of $u_{\rm max}$ as described above, see detailed description of $u$ in section 3 in the Supplemental Material [\onlinecite{SI}]) resulting in a $De$ in the range 1.4~\textendash~400. Elastic effects are thus important.

These dimensionless numbers are calculated using estimates of the maximum flow rates between pillars and using the bulk properties of the DNA solution and should be considered as nominal values and guides for comparisons only since the presence of vortices and the variation in the local concentration of DNA (as easily seen in Fig.~\ref{fig:high_res_micrographs_and_kymographs}) leads to large variations in these numbers in time and space. What is more, the measured flow rates for the given applied driving pressures differ between the flatwise and pointwise directions, leading to different $De$ (see discussion on diodicity below). 

While an array of circular pillars generates two types of waves with orientations mirrored around the axis along the device channel with equal probability, as previously reported [\onlinecite{strom_waves2022}], only one type of wave dominates for each flow direction when we break the lateral symmetry using the triangular pillars. Not only do the waves differ in their angle, see Fig.~\ref{fig:snapshots_array}, but there are other striking differences. In both directions, waves are preceded by an increase in fluctuations, see Fig.~\ref{fig:high_res_micrographs_and_kymographs}. In the flatwise direction vortex pairs form that appear to stabilize the flow between waves, which in turn may be the reason that waves appear at higher flow rates for the flatwise flow than for the pointwise flow.

%OS: Figure size: 1 column width (86 mm), 1.5 column widths (129 mm) or 2 column widths (172 mm)

Analyzing the spatial, two-dimensional (2D) frequency components of the macroscopic flow patterns seen in Fig.~\ref{fig:snapshots_array} reveals a distinct difference in the Fourier transform amplitude spectra of the two flow directions above approximately \SI{3e-3}{\m\per\s}, see Fig.~\ref{fig:fourier_spectra} for the low frequency components. The angles corresponding to a high amplitude for each flow direction in the high velocity spectra (\SI{3e-3}{\m\per\s}, \SI{6e-3}{\m\per\s}, and \SI{1e-2}{\m\per\s}) correspond to the orientations of the waves of the two directions visible in Fig.~\ref{fig:snapshots_array}. It is again clear that the waves corresponding to the pointwise flow (green) appear for lower flow velocities than the waves corresponding to the flatwise flow (red). See Supplemental Material Figs. S7 and S8 [\onlinecite{SI}] for amplitude spectra of the full frequency range as well as a large number of flow velocities.

On the microscopic scale of the pillars, for flow in both directions, as the flow velocity is increased from zero, the first observation we make is that stable deadzones depleted of DNA form and that their distributions differ between the two directions, Fig.~\ref{fig:high_res_micrographs_and_kymographs}(a and g). This corresponds to what has been reported in viscoelastic flows in pillar arrays of different shapes [\onlinecite{kawale2017, kawale2019}], and is similar to what we see for circular pillars [\onlinecite{strom_waves2022}]. As the flow velocity is increased, the asymmetry becomes more pronounced. In both directions vortices form but their positions, strength, fluctuations, and interactions with each other are different, see Fig.~\ref{fig:high_res_micrographs_and_kymographs}(b-d and h). Finally, as can be seen in Fig.~\ref{fig:snapshots_array} and Fig.~\ref{fig:high_res_micrographs_and_kymographs}(e, f and i-l) waves form that are very different on both the microscopic and macroscopic scale.

Flow in the flatwise direction is dominated by the formation of non-symmetric vortex pairs between pillars along the flow direction, with a smaller clock-wise vortex near the vertex of the pillar and a larger counter clock-wise brighter vortex below. The formation of these vortex pairs can be clearly seen in Fig.~\ref{fig:high_res_micrographs_and_kymographs}(d) at $u =$ \SI{4.7}{\milli\meter\per\second} and they persist between waves at higher flow rates, Fig.~\ref{fig:high_res_micrographs_and_kymographs}(e and f). Immediately prior to the passage of a wave the vortices appear to entangle and subsequently mix and vanish. The waves are associated with a fairly abrupt accumulation of DNA in the vortices as is clear from the insets in Fig.~\ref{fig:high_res_micrographs_and_kymographs}(f). For a better view of the dynamics we refer to the Supplemental Movies S2 and S3 [\onlinecite{SI}].

In the pointwise direction, the evolution of the flow pattern is not as clear cut as for the flatwise flow. Stable vortices are not observed. Instead, the flow of DNA exhibits two distinct phenomena as laid out in Fig.~\ref{fig:high_res_time_signal_analysis}(g and h). In (h) rather stable, yet minuscule wakes form at each downstream vertex during the period between waves resembling those described, for example, in reference [\onlinecite{HawardShenCylinder2020}]. In (g) the wakes vanish as the wave of high-concentration DNA passes. In contrast to the flatwise direction, here waves are associated with a gradual shift of the accumulated DNA as is seen in the inset in Fig.~\ref{fig:high_res_micrographs_and_kymographs}(l). For a better view of the dynamics we refer to the Supplemental Movies S2 and S3 [\onlinecite{SI}].

An analysis of the intensity in a unit cell of the array around one pillar further highlights the differences in flow behaviour in the two flow directions. Fig.~\ref{fig:high_res_time_signal_analysis}(a and c) shows clear difference of the time evolution of the intensities around a pillar due to the propagation of the wavefronts for the two flow directions in terms of the regularity of the fluctuations. Moreover,the detailed shapes of the individual peaks differ as is seen in Fig.~\ref{fig:high_res_time_signal_analysis}(b and d). A Fourier analysis of the time signals (see the Supplemental Material [\onlinecite{SI}], Figs. S4 and S5) from a unit cell [as defined in Fig.~\ref{fig:high_res_time_signal_analysis}(g)] identifies a clear frequency peak for flatwise flow as opposed to what is seen for pointwise flow.

We observe a rapid increase in the measured flow rate at the onset of waves. While the measured flow rate between the two flow directions for the same applied pressure [see \textit{e.g.}, Fig.~\ref{fig:high_res_time_signal_analysis}(e)] is larger for the pointwise flow than for the flatwise flow across the range of driving pressures observed, the difference is at a maximum in the range where waves occur for the pointwise flow, but not for the flatwise flow. We ascribe this measured difference to the difference in the onset of waves in the two directions. Movies S4 and S5 in the Supplemental Material [\onlinecite{SI}] shows this difference clearly during the application of an oscillating driving pressure.

The work constitutes a first step to detailed understanding of the wave phenomenon. However, for detailed understanding of the underlying mechanisms, careful high-speed PIV measurements will be necessary, especially during the passage of a wave across a pillar. Understanding the waves will aid the design of effective low $Re$ mixers and enable optimization of devices that use pillar arrays for the separation of DNA, for example, in separation systems with symmetric pillars that have been used for the separation of DNA where waves are observed to be equally prevalent in two symmetric orientations leading to considerable dispersion of molecular fractions [\onlinecite{strom2022}]. We propose that the ability to control the nature and directions of waves will allow improved DNA separations at high concentrations and high throughputs. Lastly, the resulting diodicity, $Di$ [see Fig.~\ref{fig:high_res_time_signal_analysis}(f)] suggests the potential application of the observed flow phenomena as a fluidic rectifier [\onlinecite{Groisman2004}].

%\begin{acknowledgments}
This research was funded by the European Union, grant number 634890 (project BeyondSeq \slash Horizon2020), EuroNanoMed (NanoDiaBac), by the Swedish Research council, grant number 2016-05739 and NanoLund. All device processing was conducted within Lund Nano Lab.
%\end{acknowledgments}

%\section*{\label{sec:author_contributions}Author Contributions}
%The author contribution statements have been written according to CRediT (Contributor Roles Taxonomy, see credit.niso.org for role descriptions): Conceptualization, O.E.S., J.P.B., and J.O.T.; Data Curation, O.E.S. and J.P.B.; Formal Analysis, O.E.S. and J.P.B.; Funding Acquisition, J.O.T.; Investigation, O.E.S., and J.P.B.; Methodology, O.E.S., J.P.B., and J.O.T.; Project Administration, J.O.T.; Resources, J.O.T.; Software, O.E.S. and J.P.B.; Supervision, J.P.B., and J.O.T.; Validation, O.E.S., J.P.B., and J.O.T.; Visualization, O.E.S., J.P.B., and J.O.T.; Writing - Original Draft, O.E.S., J.P.B., and J.O.T.; Writing - Reviewing \& Editing, O.E.S., J.P.B., and J.O.T.

%OS: Include citations from the Supplemental Material, however invisibly along the following style:
%\nocite{key}

\clearpage
\nocite{SI}
\bibliography{symmetry_library}% Produces the bibliography via BibTeX.

\begin{thebibliography}{19}%
\makeatletter
\providecommand \@ifxundefined [1]{%
 \@ifx{#1\undefined}
}%
\providecommand \@ifnum [1]{%
 \ifnum #1\expandafter \@firstoftwo
 \else \expandafter \@secondoftwo
 \fi
}%
\providecommand \@ifx [1]{%
 \ifx #1\expandafter \@firstoftwo
 \else \expandafter \@secondoftwo
 \fi
}%
\providecommand \natexlab [1]{#1}%
\providecommand \enquote  [1]{``#1''}%
\providecommand \bibnamefont  [1]{#1}%
\providecommand \bibfnamefont [1]{#1}%
\providecommand \citenamefont [1]{#1}%
\providecommand \href@noop [0]{\@secondoftwo}%
\providecommand \href [0]{\begingroup \@sanitize@url \@href}%
\providecommand \@href[1]{\@@startlink{#1}\@@href}%
\providecommand \@@href[1]{\endgroup#1\@@endlink}%
\providecommand \@sanitize@url [0]{\catcode `\\12\catcode `\$12\catcode
  `\&12\catcode `\#12\catcode `\^12\catcode `\_12\catcode `\%12\relax}%
\providecommand \@@startlink[1]{}%
\providecommand \@@endlink[0]{}%
\providecommand \url  [0]{\begingroup\@sanitize@url \@url }%
\providecommand \@url [1]{\endgroup\@href {#1}{\urlprefix }}%
\providecommand \urlprefix  [0]{URL }%
\providecommand \Eprint [0]{\href }%
\providecommand \doibase [0]{https://doi.org/}%
\providecommand \selectlanguage [0]{\@gobble}%
\providecommand \bibinfo  [0]{\@secondoftwo}%
\providecommand \bibfield  [0]{\@secondoftwo}%
\providecommand \translation [1]{[#1]}%
\providecommand \BibitemOpen [0]{}%
\providecommand \bibitemStop [0]{}%
\providecommand \bibitemNoStop [0]{.\EOS\space}%
\providecommand \EOS [0]{\spacefactor3000\relax}%
\providecommand \BibitemShut  [1]{\csname bibitem#1\endcsname}%
\let\auto@bib@innerbib\@empty
%</preamble>
\bibitem [{\citenamefont {Sorbie}(1991)}]{Sorbie1991_book}%
  \BibitemOpen
  \bibfield  {author} {\bibinfo {author} {\bibfnamefont {K.~S.}\ \bibnamefont
  {Sorbie}},\ }\href@noop {} {\emph {\bibinfo {title} {Polymer-improved oil
  recovery}}}\ (\bibinfo  {publisher} {Blackie ; CRC Press},\ \bibinfo
  {address} {Glasgow Boca Raton, Fla.},\ \bibinfo {year} {1991})\ pp.\ \bibinfo
  {pages} {xii, 359 p.}\BibitemShut {Stop}%
\bibitem [{\citenamefont {Carrel}\ \emph {et~al.}(2018)\citenamefont {Carrel},
  \citenamefont {Morales}, \citenamefont {Beltran}, \citenamefont {Derlon},
  \citenamefont {Kaufmann}, \citenamefont {Morgenroth},\ and\ \citenamefont
  {Holzner}}]{Carrel2018}%
  \BibitemOpen
  \bibfield  {author} {\bibinfo {author} {\bibfnamefont {M.}~\bibnamefont
  {Carrel}}, \bibinfo {author} {\bibfnamefont {V.~L.}\ \bibnamefont {Morales}},
  \bibinfo {author} {\bibfnamefont {M.~A.}\ \bibnamefont {Beltran}}, \bibinfo
  {author} {\bibfnamefont {N.}~\bibnamefont {Derlon}}, \bibinfo {author}
  {\bibfnamefont {R.}~\bibnamefont {Kaufmann}}, \bibinfo {author}
  {\bibfnamefont {E.}~\bibnamefont {Morgenroth}},\ and\ \bibinfo {author}
  {\bibfnamefont {M.}~\bibnamefont {Holzner}},\ }\href
  {https://doi.org/10.1016/j.watres.2018.01.059} {\bibfield  {journal}
  {\bibinfo  {journal} {Water Research}\ }\textbf {\bibinfo {volume} {134}},\
  \bibinfo {pages} {280} (\bibinfo {year} {2018})}\BibitemShut {NoStop}%
\bibitem [{\citenamefont {Groisman}\ and\ \citenamefont
  {Steinberg}(2001)}]{Groisman2011}%
  \BibitemOpen
  \bibfield  {author} {\bibinfo {author} {\bibfnamefont {A.}~\bibnamefont
  {Groisman}}\ and\ \bibinfo {author} {\bibfnamefont {V.}~\bibnamefont
  {Steinberg}},\ }\href {https://doi.org/Doi 10.1038/35073524} {\bibfield
  {journal} {\bibinfo  {journal} {Nature}\ }\textbf {\bibinfo {volume} {410}},\
  \bibinfo {pages} {905} (\bibinfo {year} {2001})}\BibitemShut {NoStop}%
\bibitem [{\citenamefont {Hochstetter}\ \emph {et~al.}(2020)\citenamefont
  {Hochstetter}, \citenamefont {Vernekar}, \citenamefont {Austin},
  \citenamefont {Becker}, \citenamefont {Beech}, \citenamefont {Fedosov},
  \citenamefont {Gompper}, \citenamefont {Kim}, \citenamefont {Smith},
  \citenamefont {Stolovitzky}, \citenamefont {Tegenfeldt}, \citenamefont
  {Wunsch}, \citenamefont {Zeming}, \citenamefont {Kruger},\ and\ \citenamefont
  {Inglis}}]{Hochstetter2020}%
  \BibitemOpen
  \bibfield  {author} {\bibinfo {author} {\bibfnamefont {A.}~\bibnamefont
  {Hochstetter}}, \bibinfo {author} {\bibfnamefont {R.}~\bibnamefont
  {Vernekar}}, \bibinfo {author} {\bibfnamefont {R.~H.}\ \bibnamefont
  {Austin}}, \bibinfo {author} {\bibfnamefont {H.}~\bibnamefont {Becker}},
  \bibinfo {author} {\bibfnamefont {J.~P.}\ \bibnamefont {Beech}}, \bibinfo
  {author} {\bibfnamefont {D.~A.}\ \bibnamefont {Fedosov}}, \bibinfo {author}
  {\bibfnamefont {G.}~\bibnamefont {Gompper}}, \bibinfo {author} {\bibfnamefont
  {S.~C.}\ \bibnamefont {Kim}}, \bibinfo {author} {\bibfnamefont {J.~T.}\
  \bibnamefont {Smith}}, \bibinfo {author} {\bibfnamefont {G.}~\bibnamefont
  {Stolovitzky}}, \bibinfo {author} {\bibfnamefont {J.~O.}\ \bibnamefont
  {Tegenfeldt}}, \bibinfo {author} {\bibfnamefont {B.~H.}\ \bibnamefont
  {Wunsch}}, \bibinfo {author} {\bibfnamefont {K.~K.}\ \bibnamefont {Zeming}},
  \bibinfo {author} {\bibfnamefont {T.}~\bibnamefont {Kruger}},\ and\ \bibinfo
  {author} {\bibfnamefont {D.~W.}\ \bibnamefont {Inglis}},\ }\href
  {https://doi.org/10.1021/acsnano.0c05186} {\bibfield  {journal} {\bibinfo
  {journal} {Acs Nano}\ }\textbf {\bibinfo {volume} {14}},\ \bibinfo {pages}
  {10784} (\bibinfo {year} {2020})}\BibitemShut {NoStop}%
\bibitem [{\citenamefont {Huang}\ \emph {et~al.}(2004)\citenamefont {Huang},
  \citenamefont {Cox}, \citenamefont {Austin},\ and\ \citenamefont
  {Sturm}}]{Huang2004}%
  \BibitemOpen
  \bibfield  {author} {\bibinfo {author} {\bibfnamefont {L.~R.}\ \bibnamefont
  {Huang}}, \bibinfo {author} {\bibfnamefont {E.~C.}\ \bibnamefont {Cox}},
  \bibinfo {author} {\bibfnamefont {R.~H.}\ \bibnamefont {Austin}},\ and\
  \bibinfo {author} {\bibfnamefont {J.~C.}\ \bibnamefont {Sturm}},\ }\href
  {https://doi.org/DOI 10.1126/science.1094567} {\bibfield  {journal} {\bibinfo
   {journal} {Science}\ }\textbf {\bibinfo {volume} {304}},\ \bibinfo {pages}
  {987} (\bibinfo {year} {2004})}\BibitemShut {NoStop}%
\bibitem [{\citenamefont {Chen}\ \emph {et~al.}(2015)\citenamefont {Chen},
  \citenamefont {Abrams}, \citenamefont {Boles}, \citenamefont {Pedersen},
  \citenamefont {Flyvbjerg}, \citenamefont {Austin},\ and\ \citenamefont
  {Sturm}}]{Chen2015}%
  \BibitemOpen
  \bibfield  {author} {\bibinfo {author} {\bibfnamefont {Y.}~\bibnamefont
  {Chen}}, \bibinfo {author} {\bibfnamefont {E.~S.}\ \bibnamefont {Abrams}},
  \bibinfo {author} {\bibfnamefont {T.~C.}\ \bibnamefont {Boles}}, \bibinfo
  {author} {\bibfnamefont {J.~N.}\ \bibnamefont {Pedersen}}, \bibinfo {author}
  {\bibfnamefont {H.}~\bibnamefont {Flyvbjerg}}, \bibinfo {author}
  {\bibfnamefont {R.~H.}\ \bibnamefont {Austin}},\ and\ \bibinfo {author}
  {\bibfnamefont {J.~C.}\ \bibnamefont {Sturm}},\ }\bibfield  {journal}
  {\bibinfo  {journal} {Physical Review Letters}\ }\textbf {\bibinfo {volume}
  {114}},\ \href {https://doi.org/ARTN 198303 10.1103/PhysRevLett.114.198303}
  {ARTN 198303 10.1103/PhysRevLett.114.198303} (\bibinfo {year}
  {2015})\BibitemShut {NoStop}%
\bibitem [{\citenamefont {Wunsch}\ \emph {et~al.}(2019)\citenamefont {Wunsch},
  \citenamefont {Kim}, \citenamefont {Gifford}, \citenamefont {Astier},
  \citenamefont {Wang}, \citenamefont {Bruce}, \citenamefont {Patel},
  \citenamefont {Duch}, \citenamefont {Dawes}, \citenamefont {Stolovitzky},\
  and\ \citenamefont {Smith}}]{Wunsch2019}%
  \BibitemOpen
  \bibfield  {author} {\bibinfo {author} {\bibfnamefont {B.~H.}\ \bibnamefont
  {Wunsch}}, \bibinfo {author} {\bibfnamefont {S.~C.}\ \bibnamefont {Kim}},
  \bibinfo {author} {\bibfnamefont {S.~M.}\ \bibnamefont {Gifford}}, \bibinfo
  {author} {\bibfnamefont {Y.}~\bibnamefont {Astier}}, \bibinfo {author}
  {\bibfnamefont {C.}~\bibnamefont {Wang}}, \bibinfo {author} {\bibfnamefont
  {R.~L.}\ \bibnamefont {Bruce}}, \bibinfo {author} {\bibfnamefont {J.~V.}\
  \bibnamefont {Patel}}, \bibinfo {author} {\bibfnamefont {E.~A.}\ \bibnamefont
  {Duch}}, \bibinfo {author} {\bibfnamefont {S.}~\bibnamefont {Dawes}},
  \bibinfo {author} {\bibfnamefont {G.}~\bibnamefont {Stolovitzky}},\ and\
  \bibinfo {author} {\bibfnamefont {J.~T.}\ \bibnamefont {Smith}},\ }\href
  {https://doi.org/10.1039/c8lc01408f} {\bibfield  {journal} {\bibinfo
  {journal} {Lab on a Chip}\ }\textbf {\bibinfo {volume} {19}},\ \bibinfo
  {pages} {1567} (\bibinfo {year} {2019})}\BibitemShut {NoStop}%
\bibitem [{\citenamefont {Wunsch}\ \emph {et~al.}(2021)\citenamefont {Wunsch},
  \citenamefont {Hsieh}, \citenamefont {Kim}, \citenamefont {Pereira},
  \citenamefont {Lukashov}, \citenamefont {Scerbo}, \citenamefont {Papalia},
  \citenamefont {Duch}, \citenamefont {Stolovitzky}, \citenamefont {Gifford},\
  and\ \citenamefont {Smith}}]{Wunsch2021}%
  \BibitemOpen
  \bibfield  {author} {\bibinfo {author} {\bibfnamefont {B.~H.}\ \bibnamefont
  {Wunsch}}, \bibinfo {author} {\bibfnamefont {K.~Y.}\ \bibnamefont {Hsieh}},
  \bibinfo {author} {\bibfnamefont {S.~C.}\ \bibnamefont {Kim}}, \bibinfo
  {author} {\bibfnamefont {M.}~\bibnamefont {Pereira}}, \bibinfo {author}
  {\bibfnamefont {S.}~\bibnamefont {Lukashov}}, \bibinfo {author}
  {\bibfnamefont {C.}~\bibnamefont {Scerbo}}, \bibinfo {author} {\bibfnamefont
  {J.~M.}\ \bibnamefont {Papalia}}, \bibinfo {author} {\bibfnamefont {E.~A.}\
  \bibnamefont {Duch}}, \bibinfo {author} {\bibfnamefont {G.}~\bibnamefont
  {Stolovitzky}}, \bibinfo {author} {\bibfnamefont {S.~M.}\ \bibnamefont
  {Gifford}},\ and\ \bibinfo {author} {\bibfnamefont {J.~T.}\ \bibnamefont
  {Smith}},\ }\bibfield  {journal} {\bibinfo  {journal} {Advanced Materials
  Technologies}\ }\textbf {\bibinfo {volume} {6}},\ \href {https://doi.org/ARTN
  2001083 10.1002/admt.202001083} {ARTN 2001083 10.1002/admt.202001083}
  (\bibinfo {year} {2021})\BibitemShut {NoStop}%
\bibitem [{\citenamefont {Str{\"o}m}\ \emph {et~al.}(2022)\citenamefont
  {Str{\"o}m}, \citenamefont {Beech},\ and\ \citenamefont
  {Tegenfeldt}}]{strom2022}%
  \BibitemOpen
  \bibfield  {author} {\bibinfo {author} {\bibfnamefont {O.~E.}\ \bibnamefont
  {Str{\"o}m}}, \bibinfo {author} {\bibfnamefont {J.~P.}\ \bibnamefont
  {Beech}},\ and\ \bibinfo {author} {\bibfnamefont {J.~O.}\ \bibnamefont
  {Tegenfeldt}},\ }\href@noop {} {\bibfield  {journal} {\bibinfo  {journal}
  {Micromachines}\ }\textbf {\bibinfo {volume} {13}},\ \bibinfo {pages} {1754}
  (\bibinfo {year} {2022})}\BibitemShut {NoStop}%
\bibitem [{\citenamefont {Bakajin}\ \emph {et~al.}(2001)\citenamefont
  {Bakajin}, \citenamefont {Duke}, \citenamefont {Tegenfeldt}, \citenamefont
  {Chou}, \citenamefont {Chan}, \citenamefont {Austin},\ and\ \citenamefont
  {Cox}}]{bakajin_hex2001}%
  \BibitemOpen
  \bibfield  {author} {\bibinfo {author} {\bibfnamefont {O.}~\bibnamefont
  {Bakajin}}, \bibinfo {author} {\bibfnamefont {T.~A.~J.}\ \bibnamefont
  {Duke}}, \bibinfo {author} {\bibfnamefont {J.}~\bibnamefont {Tegenfeldt}},
  \bibinfo {author} {\bibfnamefont {C.~F.}\ \bibnamefont {Chou}}, \bibinfo
  {author} {\bibfnamefont {S.~S.}\ \bibnamefont {Chan}}, \bibinfo {author}
  {\bibfnamefont {R.~H.}\ \bibnamefont {Austin}},\ and\ \bibinfo {author}
  {\bibfnamefont {E.~C.}\ \bibnamefont {Cox}},\ }\href@noop {} {\bibfield
  {journal} {\bibinfo  {journal} {Analytical Chemistry}\ }\textbf {\bibinfo
  {volume} {73}},\ \bibinfo {pages} {6053} (\bibinfo {year}
  {2001})}\BibitemShut {NoStop}%
\bibitem [{\citenamefont {Huang}\ \emph {et~al.}(2002)\citenamefont {Huang},
  \citenamefont {Tegenfeldt}, \citenamefont {Kraeft}, \citenamefont {Sturm},
  \citenamefont {Austin},\ and\ \citenamefont {Cox}}]{huang_hex2002}%
  \BibitemOpen
  \bibfield  {author} {\bibinfo {author} {\bibfnamefont {L.~R.}\ \bibnamefont
  {Huang}}, \bibinfo {author} {\bibfnamefont {J.~O.}\ \bibnamefont
  {Tegenfeldt}}, \bibinfo {author} {\bibfnamefont {J.~J.}\ \bibnamefont
  {Kraeft}}, \bibinfo {author} {\bibfnamefont {J.~C.}\ \bibnamefont {Sturm}},
  \bibinfo {author} {\bibfnamefont {R.~H.}\ \bibnamefont {Austin}},\ and\
  \bibinfo {author} {\bibfnamefont {E.~C.}\ \bibnamefont {Cox}},\ }\href
  {https://doi.org/10.1038/nbt733} {\bibfield  {journal} {\bibinfo  {journal}
  {Nature Biotechnology}\ }\textbf {\bibinfo {volume} {20}},\ \bibinfo {pages}
  {1048} (\bibinfo {year} {2002})}\BibitemShut {NoStop}%
\bibitem [{\citenamefont {Ström}\ \emph {et~al.}(2022)\citenamefont {Ström},
  \citenamefont {Tegenfeldt},\ and\ \citenamefont {Beech}}]{strom_waves2022}%
  \BibitemOpen
  \bibfield  {author} {\bibinfo {author} {\bibfnamefont {O.~E.}\ \bibnamefont
  {Ström}}, \bibinfo {author} {\bibfnamefont {J.~O.}\ \bibnamefont
  {Tegenfeldt}},\ and\ \bibinfo {author} {\bibfnamefont {J.~P.}\ \bibnamefont
  {Beech}},\ }\bibfield  {journal} {\bibinfo  {journal} {ArXiv e-prints}\
  }\textbf {\bibinfo {volume} {arXiv:2211.16294 [physics.flu-dyn]}},\ \href
  {https://doi.org/https://doi.org/10.48550/arXiv.2211.16294}
  {https://doi.org/10.48550/arXiv.2211.16294} (\bibinfo {year}
  {2022})\BibitemShut {NoStop}%
\bibitem [{\citenamefont {Kawale}\ \emph {et~al.}(2017)\citenamefont {Kawale},
  \citenamefont {Marques}, \citenamefont {Zitha}, \citenamefont {Kreutzer},
  \citenamefont {Rossen},\ and\ \citenamefont {Boukany}}]{kawale2017}%
  \BibitemOpen
  \bibfield  {author} {\bibinfo {author} {\bibfnamefont {D.}~\bibnamefont
  {Kawale}}, \bibinfo {author} {\bibfnamefont {E.}~\bibnamefont {Marques}},
  \bibinfo {author} {\bibfnamefont {P.~L.}\ \bibnamefont {Zitha}}, \bibinfo
  {author} {\bibfnamefont {M.~T.}\ \bibnamefont {Kreutzer}}, \bibinfo {author}
  {\bibfnamefont {W.~R.}\ \bibnamefont {Rossen}},\ and\ \bibinfo {author}
  {\bibfnamefont {P.~E.}\ \bibnamefont {Boukany}},\ }\href@noop {} {\bibfield
  {journal} {\bibinfo  {journal} {Soft matter}\ }\textbf {\bibinfo {volume}
  {13}},\ \bibinfo {pages} {765} (\bibinfo {year} {2017})}\BibitemShut
  {NoStop}%
\bibitem [{\citenamefont {Kawale}\ \emph {et~al.}(2019)\citenamefont {Kawale},
  \citenamefont {Jayaraman},\ and\ \citenamefont {Boukany}}]{kawale2019}%
  \BibitemOpen
  \bibfield  {author} {\bibinfo {author} {\bibfnamefont {D.}~\bibnamefont
  {Kawale}}, \bibinfo {author} {\bibfnamefont {J.}~\bibnamefont {Jayaraman}},\
  and\ \bibinfo {author} {\bibfnamefont {P.~E.}\ \bibnamefont {Boukany}},\
  }\bibfield  {journal} {\bibinfo  {journal} {Biomicrofluidics}\ }\textbf
  {\bibinfo {volume} {13}},\ \href {https://doi.org/Artn 014111
  10.1063/1.5050201} {Artn 014111 10.1063/1.5050201} (\bibinfo {year}
  {2019})\BibitemShut {NoStop}%
\bibitem [{\citenamefont {Loutherback}\ \emph {et~al.}(2009)\citenamefont
  {Loutherback}, \citenamefont {Puchalla}, \citenamefont {Austin},\ and\
  \citenamefont {Sturm}}]{Loutherback2009}%
  \BibitemOpen
  \bibfield  {author} {\bibinfo {author} {\bibfnamefont {K.}~\bibnamefont
  {Loutherback}}, \bibinfo {author} {\bibfnamefont {J.}~\bibnamefont
  {Puchalla}}, \bibinfo {author} {\bibfnamefont {R.~H.}\ \bibnamefont
  {Austin}},\ and\ \bibinfo {author} {\bibfnamefont {J.~C.}\ \bibnamefont
  {Sturm}},\ }\href {http://link.aps.org/doi/10.1103/PhysRevLett.102.045301}
  {\bibfield  {journal} {\bibinfo  {journal} {Physical Review Letters}\
  }\textbf {\bibinfo {volume} {102}},\ \bibinfo {pages} {045301} (\bibinfo
  {year} {2009})}\BibitemShut {NoStop}%
\bibitem [{\citenamefont {Loutherback}\ \emph {et~al.}(2010)\citenamefont
  {Loutherback}, \citenamefont {Chou}, \citenamefont {Newman}, \citenamefont
  {Puchalla}, \citenamefont {Austin},\ and\ \citenamefont
  {Sturm}}]{Loutherback2010}%
  \BibitemOpen
  \bibfield  {author} {\bibinfo {author} {\bibfnamefont {K.}~\bibnamefont
  {Loutherback}}, \bibinfo {author} {\bibfnamefont {K.~S.}\ \bibnamefont
  {Chou}}, \bibinfo {author} {\bibfnamefont {J.}~\bibnamefont {Newman}},
  \bibinfo {author} {\bibfnamefont {J.}~\bibnamefont {Puchalla}}, \bibinfo
  {author} {\bibfnamefont {R.~H.}\ \bibnamefont {Austin}},\ and\ \bibinfo
  {author} {\bibfnamefont {J.~C.}\ \bibnamefont {Sturm}},\ }\href
  {https://doi.org/10.1007/s10404-010-0635-y} {\bibfield  {journal} {\bibinfo
  {journal} {Microfluidics and Nanofluidics}\ }\textbf {\bibinfo {volume}
  {9}},\ \bibinfo {pages} {1143} (\bibinfo {year} {2010})}\BibitemShut
  {NoStop}%
\bibitem [{SI()}]{SI}%
  \BibitemOpen
  \href@noop {} {}\bibinfo {note} {See Supplemental Material at [URL will be
  inserted by publisher] for details about experimental methods, data analysis,
  supplemental figures and movies}\BibitemShut {NoStop}%
\bibitem [{\citenamefont {Haward}\ \emph {et~al.}(2020)\citenamefont {Haward},
  \citenamefont {Hopkins},\ and\ \citenamefont
  {Shen}}]{HawardShenCylinder2020}%
  \BibitemOpen
  \bibfield  {author} {\bibinfo {author} {\bibfnamefont {S.~J.}\ \bibnamefont
  {Haward}}, \bibinfo {author} {\bibfnamefont {C.~C.}\ \bibnamefont
  {Hopkins}},\ and\ \bibinfo {author} {\bibfnamefont {A.~Q.}\ \bibnamefont
  {Shen}},\ }\bibfield  {journal} {\bibinfo  {journal} {Journal of
  Non-Newtonian Fluid Mechanics}\ }\textbf {\bibinfo {volume} {278}},\ \href
  {https://doi.org/10.1016/j.jnnfm.2020.104250} {10.1016/j.jnnfm.2020.104250}
  (\bibinfo {year} {2020})\BibitemShut {NoStop}%
\bibitem [{\citenamefont {Groisman}\ and\ \citenamefont
  {Quake}(2004)}]{Groisman2004}%
  \BibitemOpen
  \bibfield  {author} {\bibinfo {author} {\bibfnamefont {A.}~\bibnamefont
  {Groisman}}\ and\ \bibinfo {author} {\bibfnamefont {S.~R.}\ \bibnamefont
  {Quake}},\ }\bibfield  {journal} {\bibinfo  {journal} {Physical Review
  Letters}\ }\textbf {\bibinfo {volume} {92}},\ \href {https://doi.org/ARTN
  094501 10.1103/PhysRevLett.92.094501} {ARTN 094501
  10.1103/PhysRevLett.92.094501} (\bibinfo {year} {2004})\BibitemShut {NoStop}%
\end{thebibliography}%

\end{document}

%\preprint{APS/123-QED}

\title{Supplemental Material for\\~\\Broken Symmetries in Microfluidic Pillar Arrays are Reflected in a Flowing DNA Solution across Multiple Length Scales}% Force line breaks with \\

\author{Jason P. Beech*}
\author{Oskar E. Ström*}%
 \author{Jonas O. Tegenfeldt†}%
%\affiliation{%
% Division of Solid State Physics, Department of Physics and NanoLund, Lund University, P.O. %Box 118, 22100 Lund, Sweden}%

\date{\today}% It is always \today, today,

\maketitle
\onecolumngrid
%\tableofcontents

\section{Device Design}
\label{sec:device_design}
\begin{figure*}
\includegraphics[width=172mm]{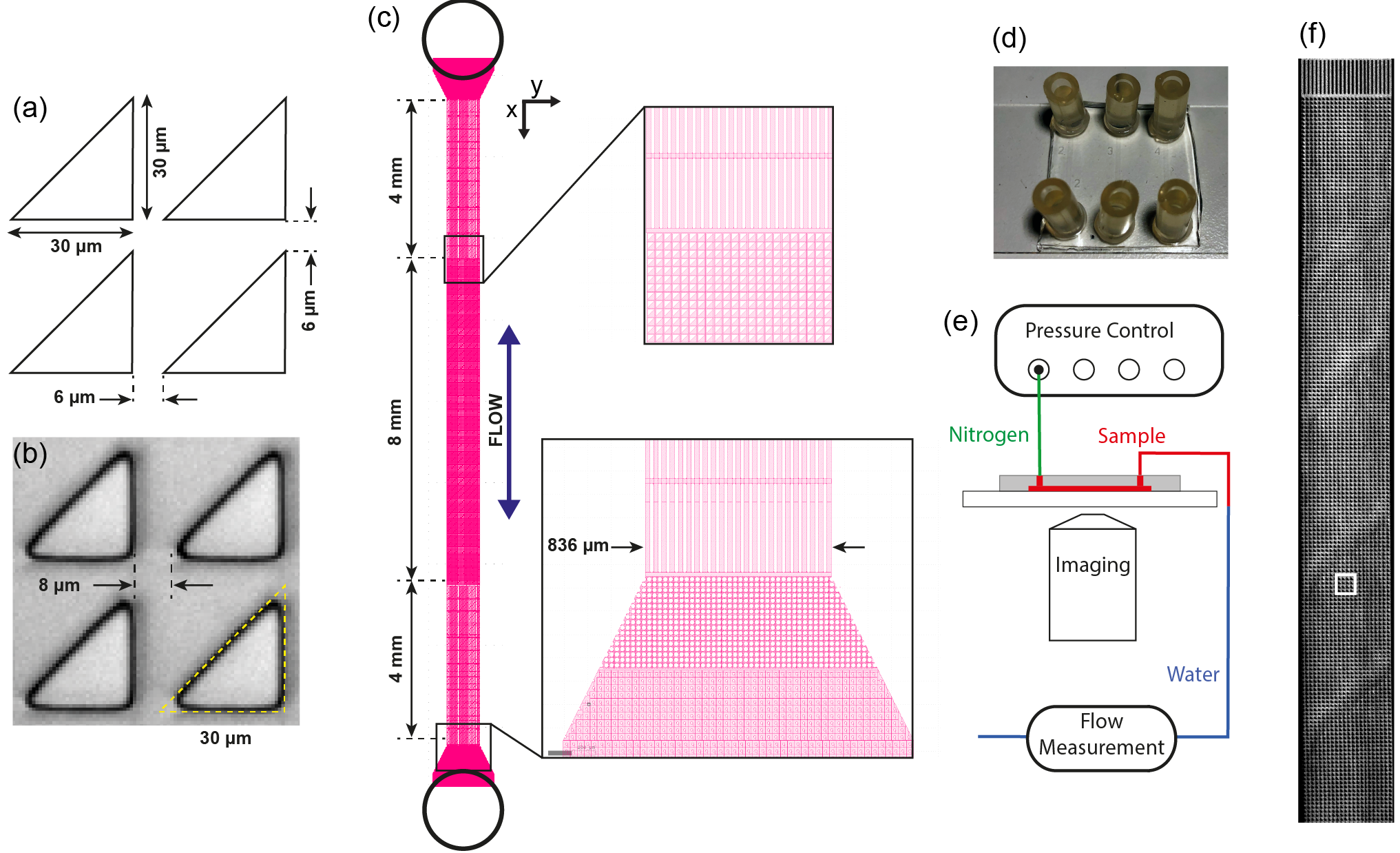}% Here is how to import EPS art
\caption{\label{fig:device_overview}Overview of the device and experimental setup. (a) Pillars were designed as right isosceles triangles. (b) Replica molded pillars in polydimethylsiloxane (PDMS). The PDMS pillars have slightly rounded corners giving a larger gap than in the design files. (c) The device consists of a channel containing an array of right-angled triangular pillars (22 pillars wide by 222 long) shown in red and straight parallel channels at either end. (d) Several devices are fabricated in PDMS in parallel for ease of experimental replication. Tubes are attached as fluid reservoirs and for connection to pressure control and flow measurement devices. (e) Flow is created in the device using pressurized nitrogen (green). The sample (red) flows through the device and is in contact with water (blue) after the outlet. Water only, for which the flow meter is calibrated, flows though the flowmeter. (f) Example fluorescent micrograph where waves can be observed. The white square indicates the approximate location where the high magnification (40$\times$~objective) micrographs presented in Fig. 3, Fig. 4(e), Fig.~\ref{fig:1D_signals}, and Fig.~\ref{fig:1D_FFT} have been captured.}
\end{figure*}

PDMS devices were cast on molds fabricated in mrDWL40 resist (Micro Resist Technology GmbH, Berlin, Germany). Four-inch wafers were treated for 60~s with oxygen plasma (Plasma Preen, NJ, USA) and resist was spun at 5000~rpm for 60~s to a thickness of 13.3~$\mu$m and exposed at approx. 450~mJ\slash cm$^2$ with a 405~nm laser using an MLA150 maskless lithography system (Heidelberg Instruments GmbH, Heidelberg, Germany). After hard baking at 120$^{\circ}$C for 10~min, the resist layer settled at 11.5 $\mu$m. A thin layer of aluminium oxide followed by a monolayer of Perfluordecyltrichlorosilane (FDTS) was deposited in an ALD system (Fiji – Plasma Enhanced ALD, Veeco, NY, USA), to assist with demolding. The cast PDMS pieces were cut to size and bonded to microscope slides via surface activation of both glass slides (60~s) and PDMS (10~s) with air plasma (Zepto, Diener electronic GmbH \& Co. KG, Ebhausen, Germany). Holes were punched through the PDMS and 3~$\times$~5~mm (inner~$\times$~outer diameter) silicone tubes of 1~cm length were glued to the PDMS to function as both liquid reservoir and connection for the application of pressurized nitrogen to drive flow (see Fig.~\ref{fig:device_overview}).

\section{\label{sec:experimental_setup}Experimental Setup}

The flow was generated by applying nitrogen gas overpressure, controlled with an MFCS-4C pressure controller (Fluigent, Paris, France). The flow rate was measured using a flow sensor (Flow rate platform with flow unit S, Fluigent, Paris, France) that was connected to the outlet tubing. The devices were imaged using an Eclipse Ti microscope (Nikon Corporation, Tokyo, Japan) and illuminated with an LED lamp (SOLA Light Engine, 6-LCR-SB, Lumencor Inc, Beaverton, OR, USA). Objectives with magnifications 2$\times$ (Nikon Plan UW, NA 0.06), 20$\times$ (Nikon CFI Plan Apochromat $\lambda$, NA 0.75), and 40$\times$ (Nikon Plan Fluor, NA 0.60) were used with a Scientific CMOS camera (Hamamatsu Orca Flash 4.0, Hamamatsu, Shizuoka Pref., Japan, pixel sizes of 6.5 $\times$ 6.5 $\mu$m). The videos were recorded with frame rates in the range 10 to \SI{33}{\per\second} depending on applied pressures.
%was used with an EMCCD camera (iXon Life 897, Andor Technology, Belfast, Northern Ireland, pixel sizes of 16 $\times$ 16 $\mu$m)
\clearpage

\section{\label{sec:calculations}Flow Calculations}
The mean flow velocity, $u$, is calculated by dividing the mean volumetric flow rate, $Q$ with the cross-sectional area of the device, $A$, as $u = Q/A$. The cross-sectional area is calculated as $A = h\cdot G \cdot N_{gaps}$, where $h =$~11.5~$\mu$m is the channel depth, $G =$~8~$\mu$m is the gap width (perpendicular to the flow) at the bottom of the triangular pillars [see Fig.~\ref{fig:device_overview}(b)], and $N_{gaps} =$~22 is the number of gaps.

Table~\ref{tab:table1} shows key parameter values for the sample used in the work. %
\begin{table}[htbp]
%The best place to locate the table environment is directly after its first reference in text
% See https://tex.stackexchange.com/questions/282884/how-to-make-a-two-column-table-entry-closer-to-the-center-of-the-table
\caption{\label{tab:table1}%
Key sample properties. The values are based on a YOYO-1 dye to base pair staining ratio of 1:200 and a 5$\times$ Tris EDTA (TE) buffer. The parameter calculations are described in [\onlinecite{strom_waves2022}]. $T = 23^{\circ}$C has been used in the calculations. $\tau$ has been measured with a stress-controlled rheometer with a 25 mm cone plate geometry at $T =25^{\circ}$C, see [\onlinecite{strom_waves2022}] for details.
}
\begin{tabular}{@{\hspace{2em}} l @{\qquad} l @{\hspace{2em}}}
\toprule
\textrm{Property}&
\textrm{Value}\\

\colrule
Number of base pairs& 48,502 \\
Polymer length, $L$& 16.61 $\mu$m \\
Molecular weight, $MW$& 31.5 MDa \\
Ionic strength, $I$& 31 mM \\
Persistence length, $l_{p}$& 51.0 nm \\
Effective width, $w_{e}$& 5.79 nm\\
Flory radius of gyration, $R_{F}$& 0.50 $\mu$m \\
Overlap concentration, $C^{*}$& 102 $\mu$g/mL \\
Measured relaxation time, $\tau$& 1.43 s \\
Zimm relaxation time, $\tau_{Zimm}$& 2.1 s \\
\botrule
\end{tabular}

\end{table}

\begin{figure*}
\includegraphics[width=178mm]{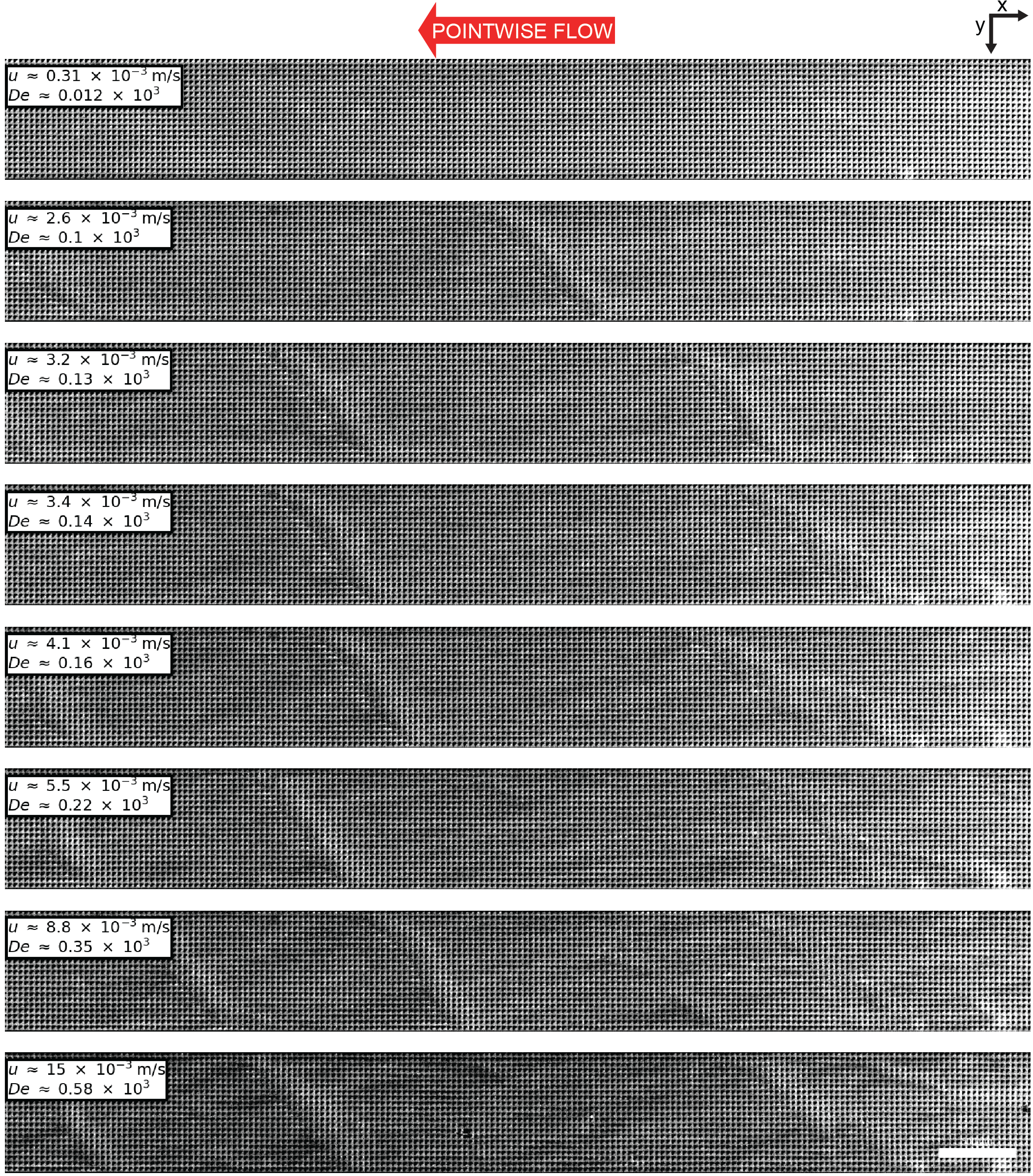}% Here is how to import EPS art
\caption{\label{fig:snapshots_2x_pointwise}Fluorescent micrograph snapshots at low magnification (2$\times$) for multiple flow velocities for pointwise DNA flow. The brightness and contrast settings have been adjusted to be the same for all micrographs. Scale bar is 500 $\mu$m.}
\end{figure*}

\begin{figure*}
\includegraphics[width=178mm]{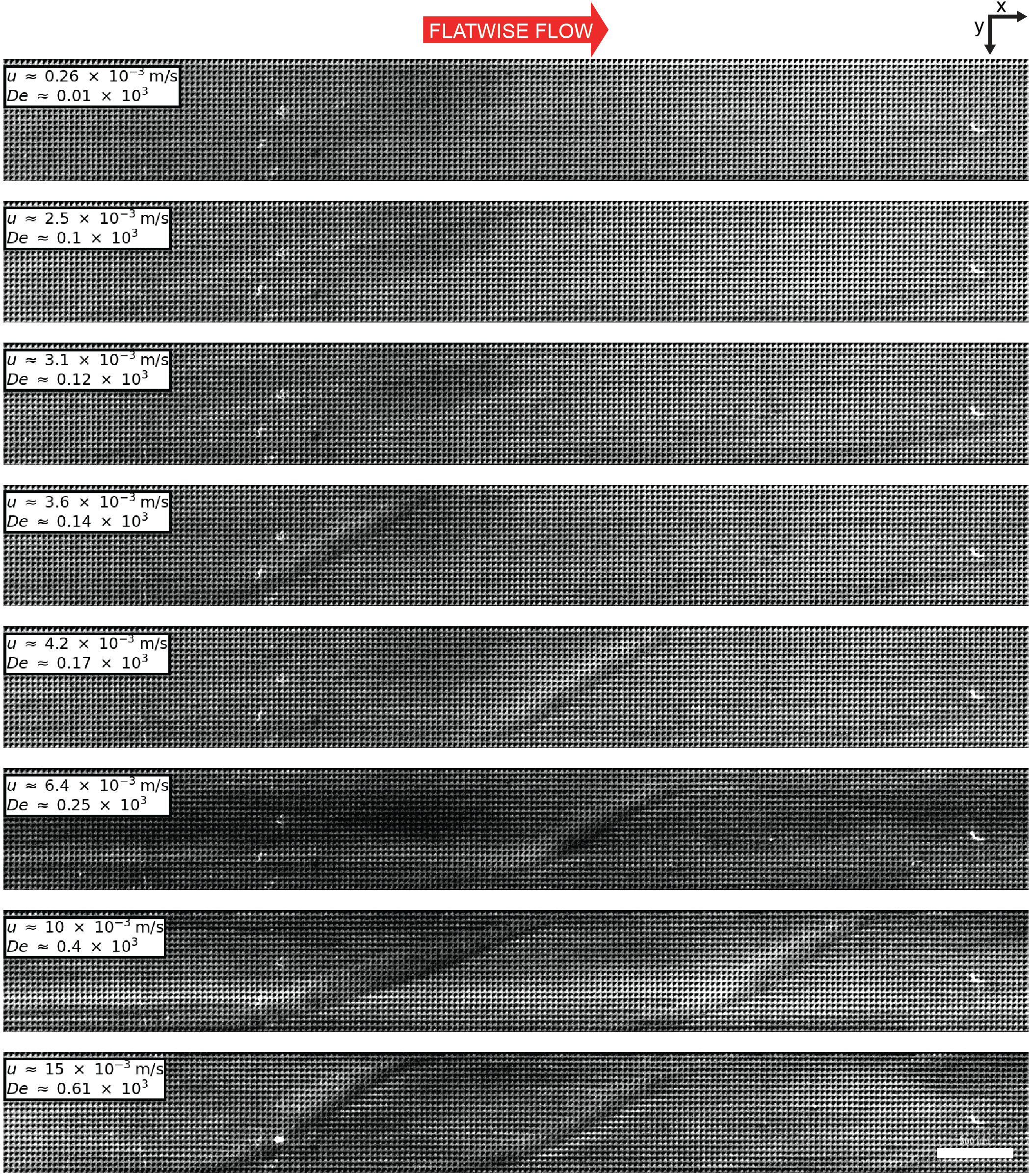}% Here is how to import EPS art
\caption{\label{fig:snapshots_2x_flatwise}Fluorescent micrograph snapshots at low magnification (2$\times$) for multiple flow velocities for flatwise DNA flow. The brightness and contrast settings have been adjusted to be the same for all micrographs. Scale bar is 500 $\mu$m.}
\end{figure*}

%\begin{figure*}
%\includegraphics[width=178mm]{ESI_figures/high_res_time_signals_many_v2-01}% Here is how to import EPS art
%\caption{\label{fig:extra_time_signals}\note{Remove this figure?}Time signals for 16 upper and lower regions for flatwise %(a) and pointwise (b) flows [see the locations of the regions (8.8 $\times$ 8.8 $\mu$m) in a unit cell in (c)]. The %location of the field of view can be seen in Fig.~\ref{fig:device_overview}(f). $u\approx$ \SI{10}{\milli\meter%\per\second}, $De \approx 400$, and $u\approx$ \SI{10}{\milli\meter\per\second}, $De \approx 350$ for flatwise and %pointwise directions, respectively. The same normalization limits have been utilized for both flow directions in (a) and %(b). Scale bar is 20 $\mu$m.}
%\end{figure*}

\clearpage
\section{\label{sec:freq_analysis}Frequency Analysis}
\subsection{\label{sec:img_pre_proc}Image Pre-processing}
A number of pre-processing steps are performed before the images are analyzed. There are defects on the camera sensor and the illumination is not fully spatially homogenous. This background is subtracted by inversely multiplying the image data with a median z-projection of a homogenous fluorescent image. The images are then rotated so that triangular pillars point rightwards and so that the array is aligned with the image axes. The average background signal of the non-fluorescent microchannel area outside of the channels is then subtracted from the image. The image is then cropped so that it is analyzed without the microchannel walls. The images are originally captured as 16-bit. To speed up the analysis, the images are converted from 16-bit to 8-bit. The conversion is performed with the same pixel value limits for the entire data set.

\subsection{\label{sec:1D_FFT}1D Frequency Analysis}
\begin{figure*}[ht]
\includegraphics[width=178mm]{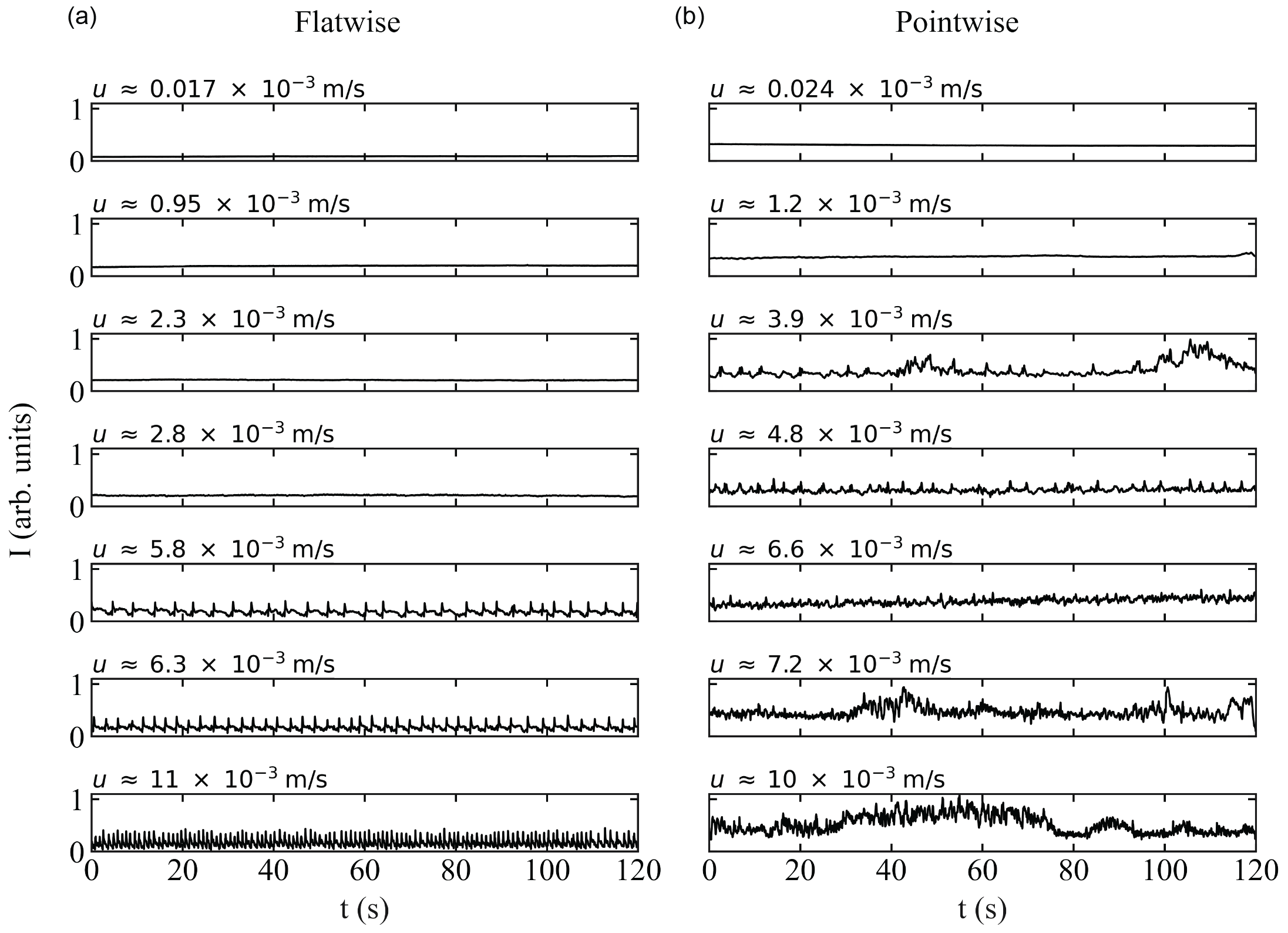}% Here is how to import EPS art
\caption{\label{fig:1D_signals}Normalized time signals of the same unit cell depending on mean flow velocity (denoted above each graph). See Fig.~4(e) for how the unit cell is spatially defined and Fig.~\ref{fig:device_overview}(f) for the location of the unit cell in the device (captured at 20$\times$ magnification).}
\end{figure*}

\begin{figure*}[ht]
\includegraphics[width=178mm]{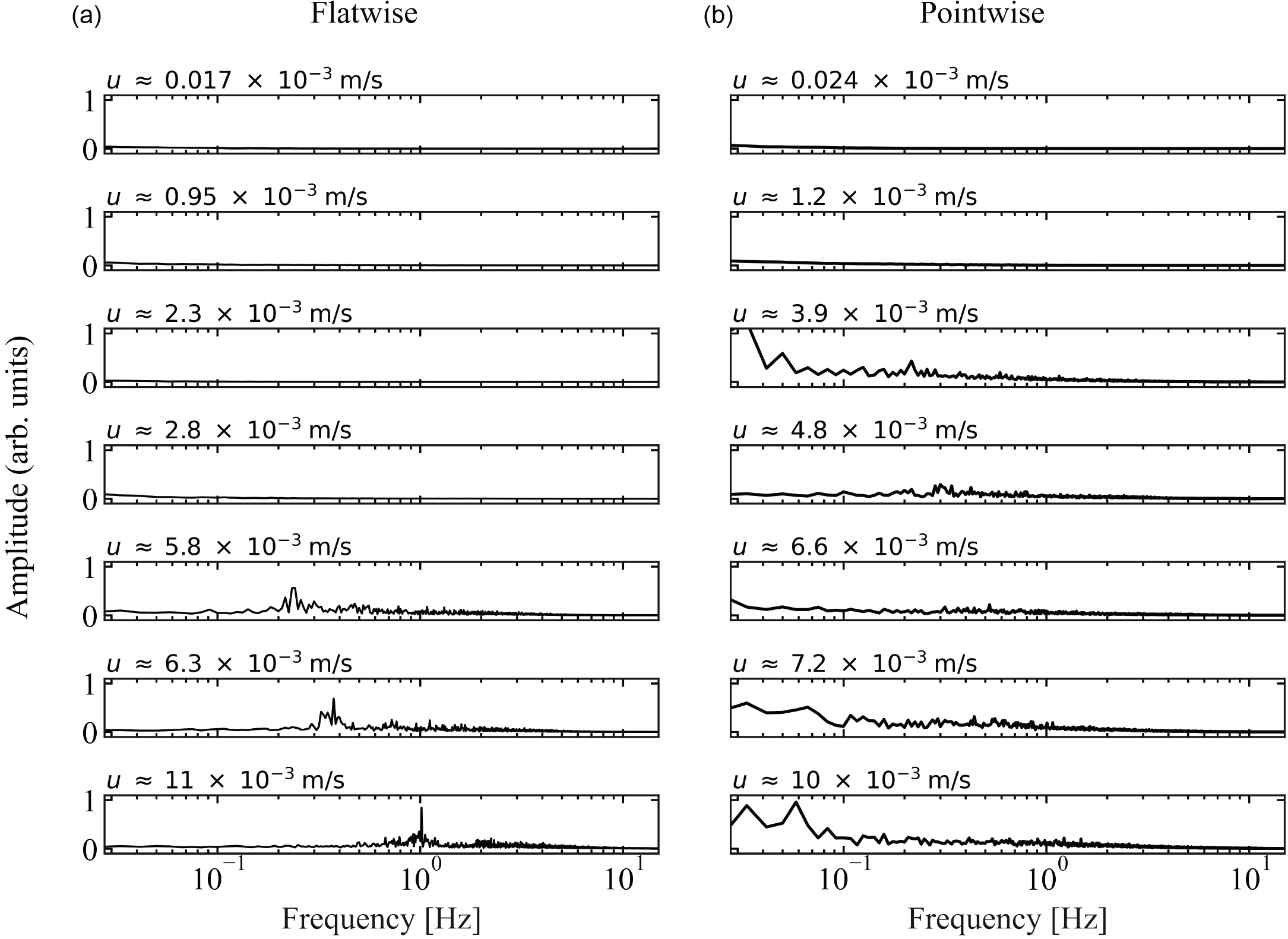}% Here is how to import EPS art
\caption{\label{fig:1D_FFT}
Normalized and mean 1D Fourier amplitude spectra of flatwise (a) and pointwise (b) flow directions for a number of mean flow velocities (denoted above each spectra). The data is based on the mean spectra of 20 unit-cell time signals (captured at 20$\times$ magnification). See examples of individual time signals in Fig.~\ref{fig:1D_signals}.}
\end{figure*}

A one-dimensional (1D) Fast Fourier Transform (FFT) function (Python Package Scipy's fft.rfft) has been applied on the mean intensity signal from 20 unit cells from either flow direction. See the intensity signals in Fig.~\ref{fig:1D_signals}. The Mean Fourier amplitude spectra are seen in  Fig.~\ref{fig:1D_FFT}.

\subsection{\label{sec:2D_FFT}2D Frequency Analysis}
\begin{figure*}[htbp]
\includegraphics[width=170mm]{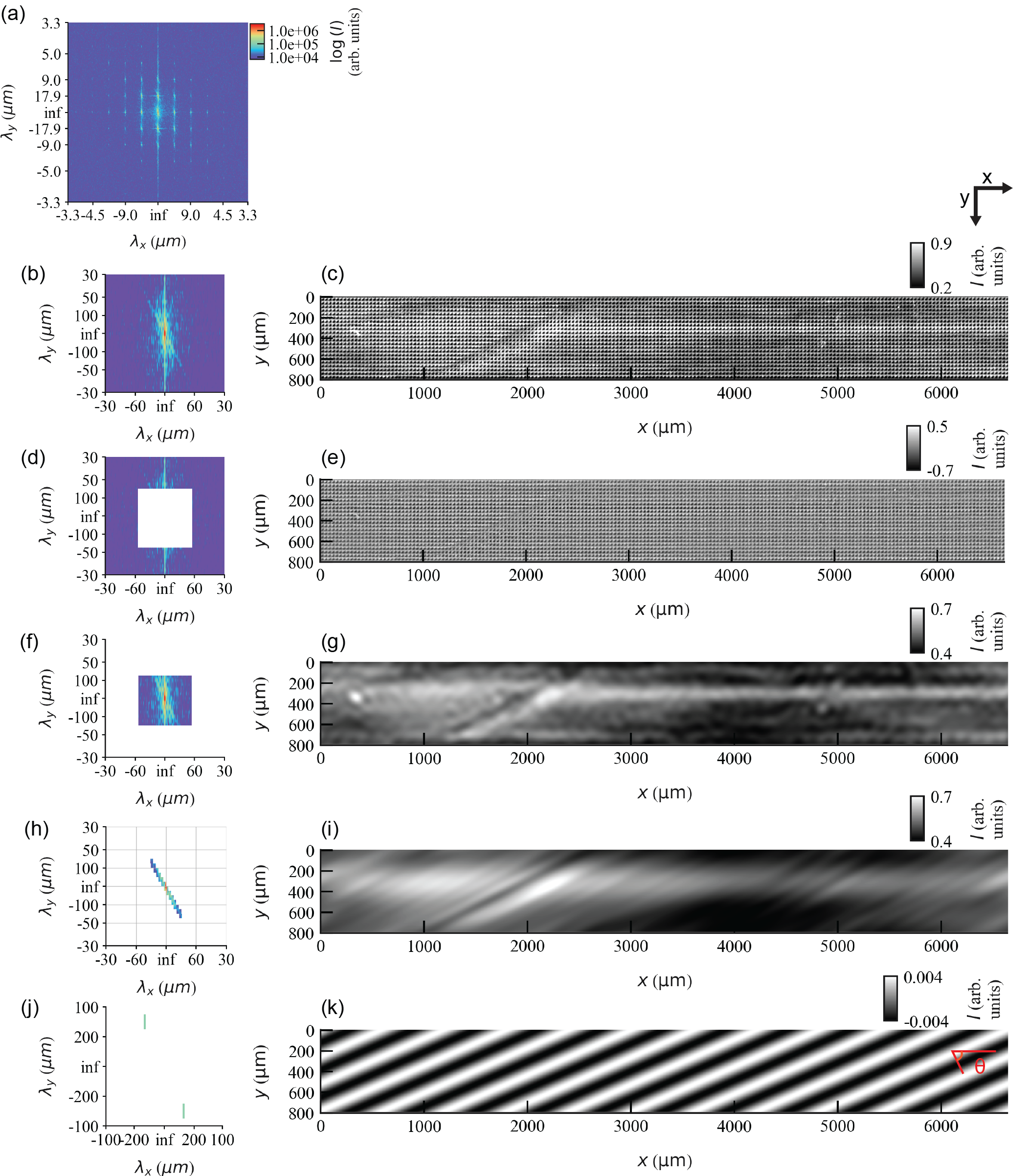}% Here is how to import EPS art
\caption{\label{fig:FFT_explanation}
Full width Fast Fourier transform (FFT) two-dimensional amplitude spectrum (a) together with frequency mask examples (b, d, f, h and j) and their corresponding inverse transforms (c, e, g, i and k), respectively. The white regions of the masks are set to zero amplitude. All spectra are based on a low magnification (2$\times$~objective) micrograph (c) with flatwise flow at $u\approx$ \SI{10}{\milli\meter\per\second} ($De \approx 400$). Both the negative and positive frequency components are shown in the spectra. (b) Low frequency spectrum [zoomed-in version of (a)]. (d) Square-shaped high-pass filter ($\lambda_{\rm x,y~cut~off}$ = 60 $\mu$m). The rectangular shape is chosen due to the image basis for the FFT is rectangular and not square which would have made it easy to generate a circular shape. (f) Square-shaped low-pass filter [inverse of (d)]. (h) Mask of the long wavelength region corresponding to the oblique line in the spectrum. (j) Mask of the frequency peak in in the long wavelength (low frequency) region with zero frequencies excluded. The peak has a wavelength of 121 $\mu$m and angle, $\theta$ = 66.5$^{\circ}$ [see drawing in (k)]. The color bar in (a) is also applicable to (b, d, f, h and j). The intensity color bar in (c, e, g, i and k) is relative to the fluorescence intensity in (c).}
\end{figure*}

Python package NumPy’s fft.fft2 function is used to create a discrete two-dimensional (2D) Fast Fourier Transform (FFT) for every frame of each video (corresponding to one pressure value or mean flow velocity) of each data set. The transform is then shifted so that the zero frequency is in the center of the two-dimensional array with Numpy’s fftshift function. See Fig.~\ref{fig:FFT_explanation} for a typical spectrum from a low magnification (2$\times$~objective) micrograph together with various masks to demonstrate what the different parts of the frequency spectrum represents in the spatial dimension.

\begin{figure*}[ht]
\includegraphics[width=178mm]{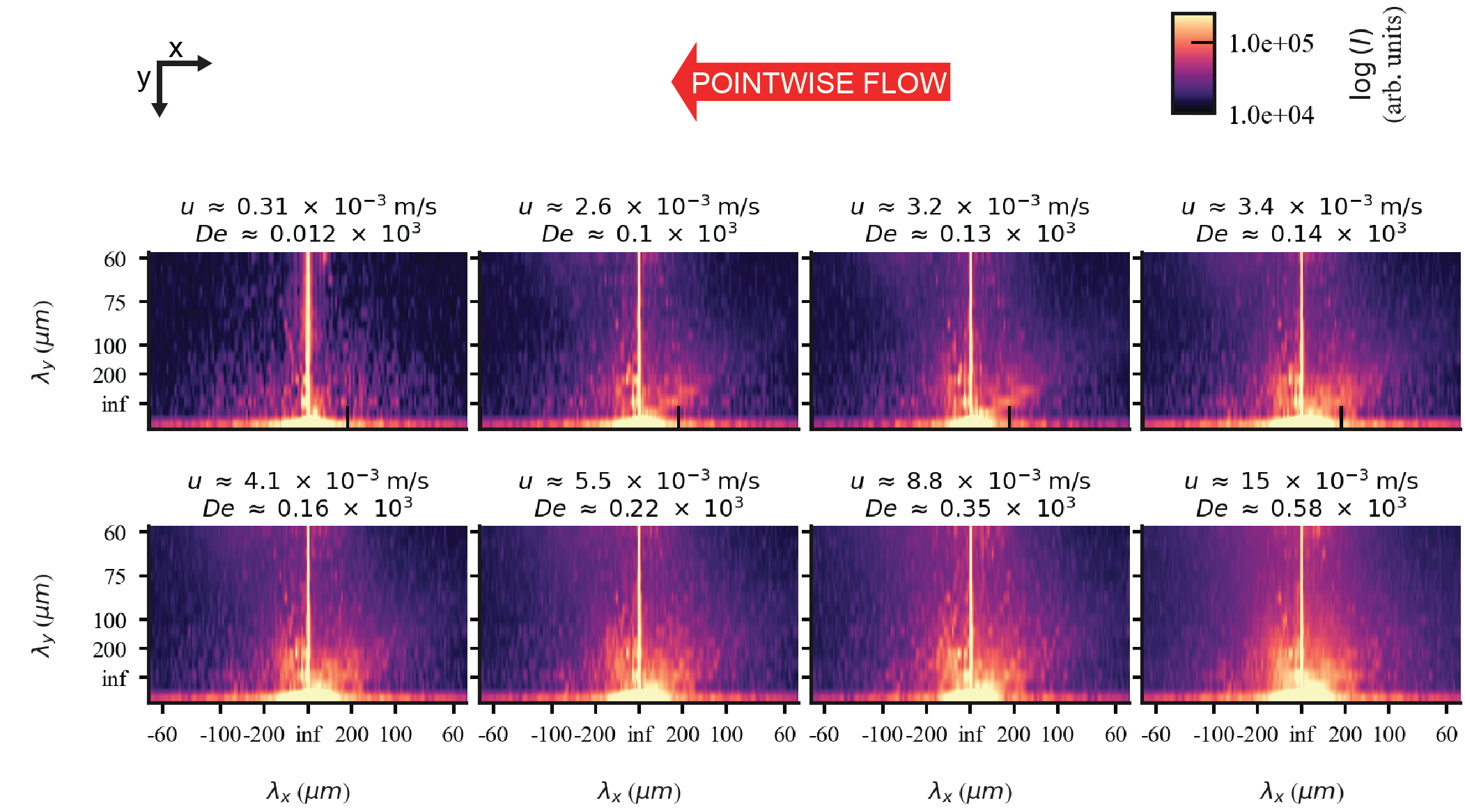}% Here is how to import EPS art
\caption{\label{fig:FFT_amplitude_spectra_pointwise}Logarithmically scaled and time-averaged two-dimensional and long wavelength (low frequency) region of Fourier amplitude spectra for pointwise flows. Only the positive y-frequencies are shown. The data is based on long fluorescence videos captured at a low magnification (2$\times$~objective) with 500, 1500, 1500, 1500, 1500, 1500, 1500, and 1500 number of frames and 1.5 min, 5 min, 5 min, 5 min, 2 min, 2 min and 2 min for the flow velocities from top to bottom, respectively. The image brightness and contrast has been set so that the absolute amplitude limits are the same for all pressure values.}
\end{figure*}

\begin{figure*}[ht]
\includegraphics[width=178mm]{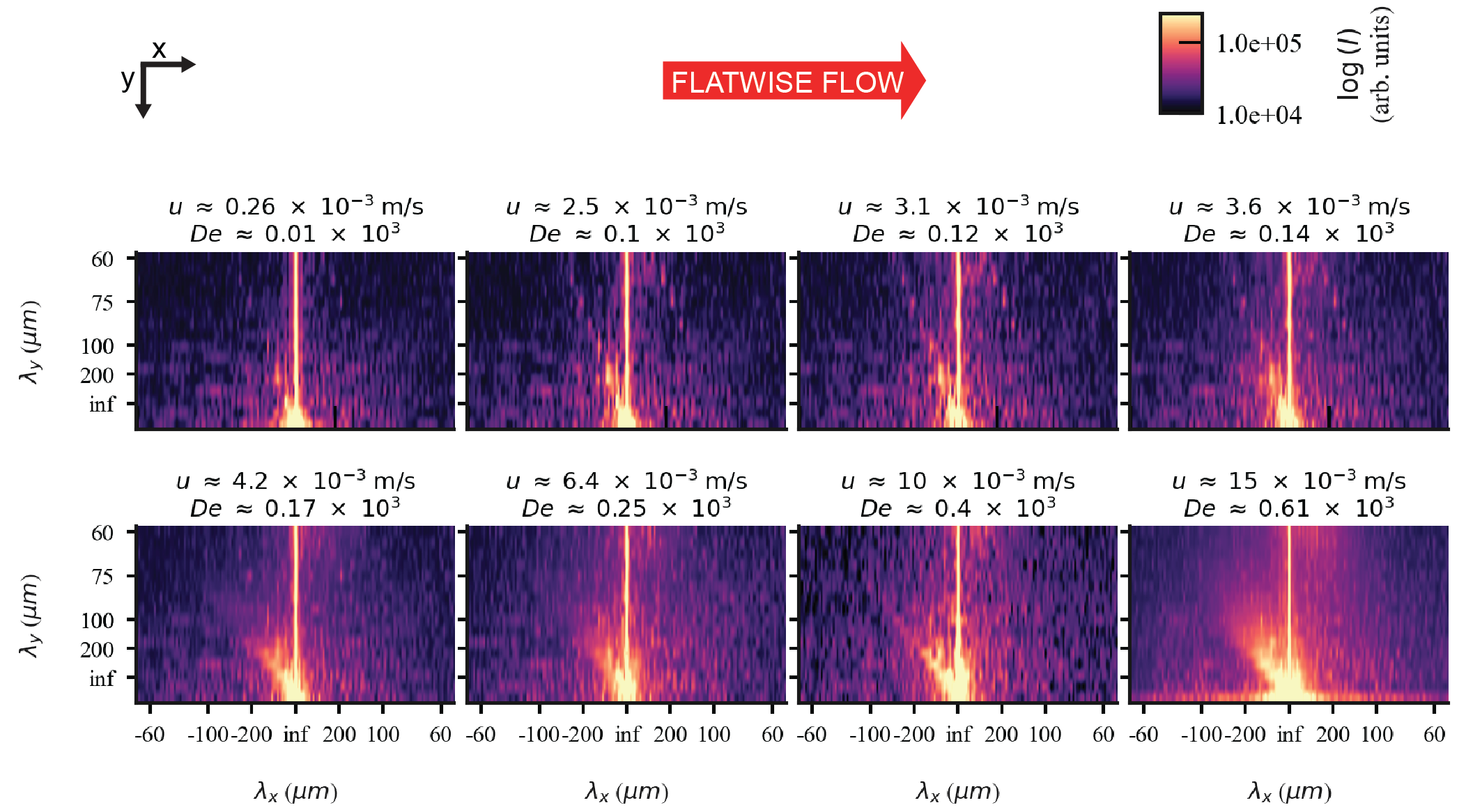}% Here is how to import EPS art
\caption{\label{fig:FFT_amplitude_spectra_flatwise}Logarithmically scaled and time-averaged two-dimensional and long wavelength (low frequency) region of Fourier amplitude spectra for flatwise flows. The data is based on long fluorescence videos captured at a low magnification (2$\times$~objective) with 500, 1500, 1500, 1500, 1500, 1500, 1500, and 1500 number of frames and 1.5 min, 5 min, 5 min, 5 min, 2 min, 2 min and 2 min for the flow velocities from top to bottom, respectively. The image brightness and contrast has been set so that the absolute amplitude limits are the same for all pressure values.}
\end{figure*}

In Fig.~2, Fig.~\ref{fig:FFT_amplitude_spectra_pointwise} and Fig.~\ref{fig:FFT_amplitude_spectra_flatwise}, the mean amplitude spectra are shown. The mean FFT spectra were computed by averging across all frames of each videograph. The logarithm of the absolute value of the amplitude are plotted. The panels show only the positive wavelengths in the y-direction (perpendicular to the flow direction) and not the superfluous negative ones. The high-magnitude at infinite wavelength corresponds to the zero frequency component.
For all 2$\times$ videographs that were analyzed by Fourier transforms, an exposure time corresponding to 50 fps was used. For the lowest flow velocity, \SI{0.3}{\milli\meter\per\second} the videographs were 500 frames or 100 s long. For flow velocities \SI{3}{\milli\meter\per\second}, \SI{4}{\milli\meter\per\second}, \SI{6}{\milli\meter\per\second}, \SI{10}{\milli\meter\per\second}, 1500 frames were used with 6 min, 2 min, 2 min and 2 min durations, respectively.
%To minimize the collection of an excessive amount of data, the number of frames for each pressure value have been optimized so that approximately 100 waves used as a base for the frequency analysis. 

\clearpage
\section{\label{sec:optical_effect}Optical Artefacts}
\begin{figure*}[!htbp]
\includegraphics[width=172mm]{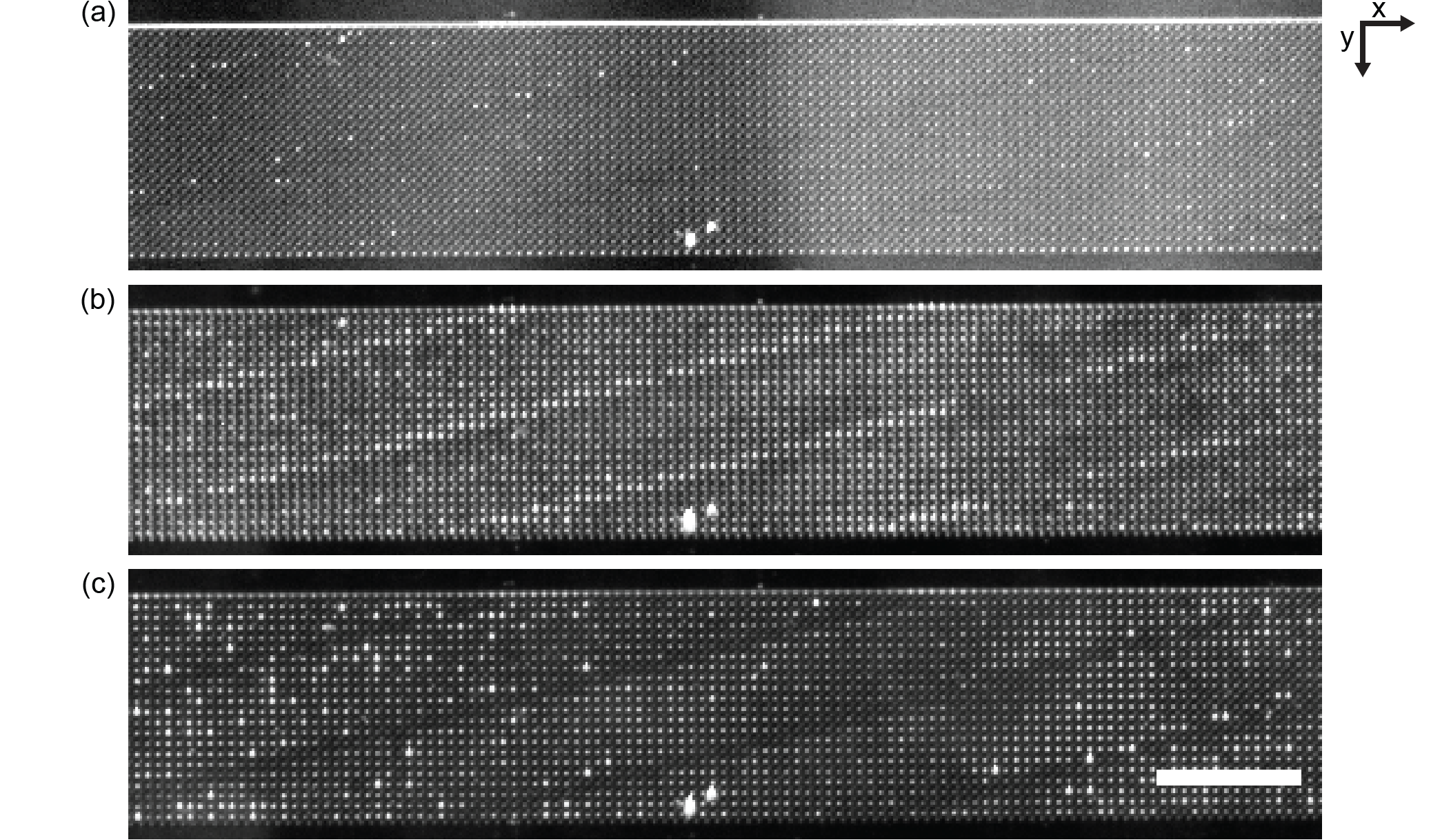}% Here is how to import EPS art
\caption{\label{fig:optical_artifacts}Brightfield micrographs array lacking any sample. (a) Illumination from top-side of the array. (b and c) Illumination from two separate side-way directions. Optical artefacts are visible as either positive (b) or negative (c) oblique stripes. Scale bar is 500 $\mu$m.}
\end{figure*}

Optical artefacts in the form of bright stripes are visible in some of the (2$\times$~objective) micrographs. For example, see the low-flow-velocity snapshots in Fig.~\ref{fig:snapshots_2x_flatwise}. At first glance, it may seem that these stripes are a product of elastic turbulence, like the waves. However, in contrast to the waves, the pattern does not move, and we observe these stripes even when the PDMS devices are empty, i.e. without a sample that could cause elastic turbulence, see Fig.~\ref{fig:optical_artifacts}. Depending on the angle of incidence of the illuminating light, these stripes vary in intensity and can either be positive or negative as seen in Fig.~\ref{fig:optical_artifacts}. We observe these stripes to always have the same angle and wavelength relative to the long axis of the array.

\clearpage
\section{Captions for Supplemental Movies}
\label{sec:captions_ESI_movies}

\noindent \textbf{Supplemental Movie S1 (\href{https://doi.org/10.5446/60253}{https://doi.org/10.5446/60253})}. Low-magnification (2$\times$ objective) fluorescent videographs at high flow velocity for pointwise ($u\approx$ \SI{8.7}{\milli\meter\per\second}, $De \approx 350$) and flatwise ($u\approx$ \SI{10}{\milli\meter\per\second}, $De \approx 400$) flows. The movie corresponds to the data shown in Fig.~1. 
\vspace{5mm} %5mm vertical space

\noindent \textbf{Supplemental Movie S2 (\href{https://doi.org/10.5446/60254}{https://doi.org/10.5446/60254})}. Two high-magnification (40$\times$ objective) fluorescent videographs side by side at a range of flow velocities depicting pointwise and flatwise flows respectively. The movie corresponds to the data shown in Fig.~4.
\vspace{5mm} %5mm vertical space

\noindent \textbf{Supplemental Movie S3 (\href{https://doi.org/10.5446/60255}{https://doi.org/10.5446/60255})}. Slowed-down high-magnification (40$\times$ objective) fluorescent videographs at high flow velocity, first for pointwise flow, followed by flatwise flows. The movie corresponds to the data shown in Fig.~4.
\vspace{5mm} %5mm vertical space

\noindent \textbf{Supplemental Movie S4 (\href{https://doi.org/10.5446/60256}{https://doi.org/10.5446/60256})}. Low-magnification (2$\times$ objective) fluorescent videograph where the applied pressure difference is oscillating from $-200$ mbar to 200 mbar. The peak flow velocity corresponds to approximately 2\textendash 4 mm/s. The size of the white bar indicates the relative pressure difference.
\vspace{5mm} %5mm vertical space

\noindent \textbf{Supplemental Movie S5 (\href{https://doi.org/10.5446/60257}{https://doi.org/10.5446/60257})}. Low-magnification (2$\times$ objective) fluorescent videograph where the applied pressure difference is oscillating from $-500$ mbar to 500 mbar. The peak flow velocity corresponds to approximately 10 mm/s. The size of the white bar indicates the relative pressure difference.

%\usepackage{caption}
%\usepackage{subcaption} 
 
%\begin{figure*}
%    \centering
%    \begin{subfigure}[b]{0.3\textwidth}
%        \centering
%        \includegraphics[width=\textwidth]{ESI_figures/amp_storage_and_loss_full_plot}
%        \caption{Amplitude sweep showing the LVE regime for 400 ng/µl lambda DNA}
%        \label{amp}
%    \end{subfigure}
    \hfill
%    \begin{subfigure}[b]{0.3\textwidth}
%        \centering
%        \includegraphics[width=\textwidth]{ESI_figures/storage_and_loss_full_plot}
%        \caption{Frequency sweep showing three independent measurements of G' and G'' for 400 ng/µl lambda DNA}
%        \label{G' G''}
%%    \hfill
%    \begin{subfigure}[b]{0.3\textwidth}
%        \centering
%        \includegraphics[width=\textwidth]{ESI_figures/g_and_gg_mean_zoom_plot}
%        \caption{The means of 3 frequency sweeps. The cross-over frequency where G’ = G’’ is shown with the black, dashed line.}
%        \label{relax time}
%\end{figure*}

\bibliography{symmetry_library}% Produces the bibliography via BibTeX.